\documentclass[amsfonts,showpacs,tightenlines,prc,nofootinbib,12pt]{revtex4}  

\usepackage{bm} 
\usepackage{epsfig}

\newcommand{\hbo}{\hbar \omega}
\newcommand{\om}{\omega}
\newcommand{\la}{\lambda}
\newcommand{\s}{\sigma}
\newcommand{\lm}{(\lambda,\mu)}
\newcommand{\lms}{(\lambda_{\sigma},\mu_{\sigma})}
\newcommand{\lmw}{(\lambda_{\omega},\mu_{\omega})}
\newcommand{\lmp}{(\lambda',\mu')}
\newcommand{\lmi}{(\lambda_i, \mu_i)}
\newcommand{\lmf}{(\lambda_f, \mu_f)}
\newcommand{\bea}{\begin{eqnarray}}
\newcommand{\eea}{ \end{eqnarray}}
\newcommand{\nn}{\nonumber} 
\newcommand{\lme}{(\lambda_1, \mu_1)}
\newcommand{\lmz}{(\lambda_2, \mu_2)}   
\newcommand{\lmd}{(\lambda_3, \mu_3)} 
\newcommand{\al}{\alpha}
\renewcommand{\k}{\kappa}
\newcommand{\m}{\mu}    
\newcommand{\cg}[6]{\langle \,#1 #2,\, #3 #4 \, | \, #5 #6 \,\rangle}
\newcommand{\sutcc}[3]{\langle \, #1 ;\, #2\, |\, #3 \,\rangle}
\newcommand{\sutrcc}[3]{\langle \, #1 ;\, #2\, ||\, #3 \,\rangle}

\begin{document}

\title{Partial Dynamical Symmetry in the Symplectic Shell Model}
\author{Jutta Escher}\email{escher@triumf.ca}
\affiliation{TRIUMF, 4004 Wesbrook Mall, Vancouver, B.C.\/ V6T 2A3,
Canada}
\author{Amiram Leviatan}\email{ami@vms.huji.ac.il}
\affiliation{Racah Institute of Physics, The Hebrew University,
Jerusalem 91904, Israel}
\date{\today}

\begin{abstract}
We present an example of a partial dynamical symmetry 
(PDS) in an interacting fermion system and demonstrate 
the close relationship of the associated Hamiltonians with a 
realistic quadrupole-quadrupole interaction, thus shedding 
new light on this important interaction.  Specifically, in 
the framework of the symplectic shell model of nuclei, we 
prove the existence of a family of fermionic Hamiltonians 
with partial SU(3) symmetry.  
We outline the construction process for the PDS eigenstates 
with good symmetry and give analytic expressions for the energies
of these states and E2 transition strengths between them.
Characteristics of both pure and mixed-symmetry PDS eigenstates 
are discussed and the resulting spectra and transition strengths 
are compared to those of real nuclei.  The PDS concept is shown 
to be relevant to the description of prolate, oblate, as well 
as triaxially deformed nuclei.  Similarities and differences 
between the fermion case and the previously established partial 
SU(3) symmetry in the Interacting Boson Model are considered.                  
\end{abstract}

\maketitle

\section{Introduction}
\label{Sec:Intro}

Symmetries play an important role in physics.
Constants of motion associated with a symmetry govern the integrability of a 
given classical system, and at the quantum level symmetries provide labels 
for the classification of states, determine selection rules, and simplify the 
relevant Hamiltonian matrices.
Algebraic, symmetry-based, theories have been firmly established as an elegant 
and practical approach to a variety of physical systems (see, for example,
Refs.~\cite{ElliottBook79,BohmEtAl88,Hamermesh,FuchsSchweigert97,
GroupsNParticles,IBMBooks,HechtSymp92,CastenBook93,TalmiBook93,Rowe96,
ELAF96,Molecules}). 
These theories offer the greatest simplifications when the interaction under 
consideration is symmetry-preserving in the selected state labeling scheme, 
that is, when the Hamiltonian either commutes with all the generators of a 
particular group (`exact symmetry') or when it is written in terms of and 
commutes with the Casimir operators of a chain of nested groups (`dynamical 
symmetry').
In both cases basis states belonging to inequivalent irreducible 
representations of the relevant groups do not mix, the Hamiltonian matrix has 
block structure, and all properties of the system can be expressed in closed 
form.
An exact or dynamical symmetry not only facilitates the numerical treatment 
of the Hamiltonian, but also its interpretation and thus provides considerable 
insight into the physics of a given system.

Naturally, the application of exact or dynamical symmetries to realistic 
situations has its limitations:
Usually the assumed symmetry is only approximately fulfilled, and imposing 
certain symmetry requirements on the Hamiltonian might result in constraints 
which are too severe and incompatible with experimentally observed features 
of the system.
The standard approach in such situations is to break the symmetry.
In cases where a symmetry-breaking Hamiltonian is involved, it is 
possible to decompose the offending terms into basic parts (``irreducible 
tensor operators'') which exhibit specific transformation properties.
Provided the appropriate group coupling coefficients and the matrix elements
of some elementary tensor operators are available, matrix elements of operators
which connect inequivalent irreducible representations can be determined and the 
exact eigenvalues and eigenstates can then be obtained (at least in principle).
While group theoretical considerations still play an important role in 
evaluating the coupling coefficients and matrix elements for such a calculation
and in truncating model spaces which have become too large for a complete 
numeric treatment, the basic simplicity of the symmetry-based approach is lost.

Alternatively, one might consider some intermediate structure, which allows 
for symmetry breaking, but preserves the advantages of a dynamical symmetry 
for a part of the system.
Partial dynamical symmetry (PDS)~\cite{GenPDS} provides such a structure. 
It corresponds to a particular symmetry breaking for which the Hamiltonian is 
not invariant under the symmetry group and hence various irreducible representions 
(irreps) are mixed in its eigenstates, yet it possesses a subset of `special' 
solvable states which respect the symmetry.
The notion of partial dynamical symmetry generalizes the concepts of exact and 
dynamical symmmetries.
In making the transition from an exact to a dynamical symmetry, states which 
are degenerate in the former scheme are split but not mixed in the latter,
and the block structure of the Hamiltonian is retained.
Proceeding further to partial symmetry, some blocks or selected states in a 
block remain pure, while other states mix and lose the symmetry character. 

Other generalizations of the idea of dynamical symmetry are possible.
Van Isacker~\cite{Piet99}, for example, suggested to break the dynamical
symmetry associated with an intermediate group $G_2$ in a subchain
$G_1 \supset G_2 \supset G_3$ for all states of the system, while 
preserving the remaining (dynamical) symmetries.
The resulting Hamiltonian is in general not analytically solvable, but its
eigenstates can still be (partly) classified by quantum labels associated
with the groups $G_1$ and $G_3$.
An approximate-symmetry scheme called quasi-dynamical symmetry was discussed
by Bahri and Rowe~\cite{QuasiDynS}.
They considered strong but coherent mixing of the irreducible representations
associated with a given dynamical symmetry.
Both methods of extending the concept of dynamical symmetry differ from the
notion of partial dynamical symmetry introduced above since, unlike in the
partial-symmetry case, the eigenvalues of the Hamiltonians cannot be
obtained analytically, not even for a part of the system.

The partial-symmetry scheme was introduced first in bosonic systems, where it was 
applied to the spectroscopy of deformed nuclei.
In Ref.~\cite{Leviatan96a}, a Hamiltonian with partial SU(3) symmetry was 
constructed in the framework of the Interacting Boson Model (IBM) of nuclei
\cite{IBMBooks}, and the calculated spectrum and E2 rates of ${}^{168}$Er 
were compared to experimental results.
The PDS Hamiltonian was found to reproduce the experimentally observed
feature of non-degenerate rotational $\gamma$ and $\beta$ bands (`K-band
splitting') and to possess several bands of solvable states, whereas
previous attempts to describe the ${}^{168}$Er data had involved 
Hamiltonians with SU(3) dynamical symmetry, which can only yield $\gamma$ 
and $\beta$ bands with degenerate angular momentum states, or had achieved
agreement with the data by completely breaking the SU(3) symmetry.
Employing the same Hamiltonian, Sinai and Leviatan~\cite{Sinai99a,Sinai99b} 
investigated the structure of the lowest collective K=$0^+$ excitation in 
deformed rare-earth nuclei.
Implications of the partial dynamical symmetry for the mixing behavior of 
this collective band were discussed and compared to broken-SU(3) predictions.
In another study, Ref.~\cite{Talmi97}, in the context of the IBM-2, the proton-neutron 
version of the Interacting Boson Model~\cite{IBMBooks,IBM2}, Talmi was able to 
explain simple regularities in spectra of the Majorana operator as an example 
of partial dynamical symmetry.
More recently, the relevance of partial F-spin symmetry was studied in the
framework of the IBM-2.
It has long been known that F spin, the SU(2) quantum number associated with
the two-valued proton-neutron degree of freedom of the IBM-2, cannot be conserved
in nuclear spectroscopy.
However, Leviatan and Ginocchio~\cite{Leviatan00a} demonstrated that empirical 
energy systematics in the deformed Dy-Os region can be reproduced under the 
assumption of partial F-spin symmetry.
Moreover, the associated partial symmetry Hamiltonians point to the existence
of F-spin multiplets of scissors states, with a moment of inertia equal to that
of the ground band.  
These predictions were tested against recent analyses of M1 transition strengths.

The subject of partial symmetries and supersymmetry in nuclear physics was
considered by Jolos and von Brentano in the context of the Interacting
Boson-Fermion Model~\cite{JolosBrentano1} and the particle-rotor 
model~\cite{JolosBrentano2}.
                                                       
Partial symmetries can be associated with continuous as well as discrete groups.
The dynamical groups employed in the IBM, e.g., are continuous.
In Ref.~\cite{PingChen97}, an example of a partial symmetry which involves point
groups was presented in the context of molecular physics.
Ping and Chen used a model of N coupled anharmonic oscillators to describe the
molecule $XY_6$.
The partial symmetry of the Hamiltonian allowed them to derive analytic expressions
for the energies of a set of unique levels and to discuss the structure of the
associated eigenstates.
Furthermore, the numerical calculations required to obtain the energies of the
remaining (non-unique) levels were greatly simplified since the Hamiltonian
could be diagonalized in a much smaller space.

Partial symmetries have relevance not only for discrete spectroscopy but 
also for the study of stochastic properties of dynamical systems.
A generic classical or quantum-mechanical Hamiltonian exhibits mixed dynamics:\
areas of regular motion and chaotic regions coexist in phase space, and even 
when a system seems to be fully chaotic, regular states may exist.
Whelan {\em et al.}~\cite{PDSChaos} used Hamiltonians with partial dynamical 
symmetries to investigate quantum-mechanical systems which are partly regular and 
partly chaotic.
In the context of the Interacting Boson Model, it was demonstrated that partial 
symmetries impose a particular phase-space structure which leads to a suppression 
of chaos in mixed systems.
Canetta and Maino~\cite{CanettaMaino} carried out a quantum-statistical
analysis of regular and chaotic dynamic behavior in the IBM-2.
Varying the Hamiltonian parameters, they observed a nearly regular region in
parameter space --- far away from dynamical symmetry limits --- which they
linked to the existence of a partial dynamical symmetry.
Since Hamiltonians with partial symmetries are not completely integrable
and may exhibit stochastic behavior, they are an ideal tool for studying
mixed systems with coexisting regularity and chaos.

Partial symmetries are not confined to bosonic systems. 
In Ref.~\cite{Escher00PRL}, an example of a partial symmetry in an interacting 
fermion system was presented.   
A family of Hamiltonians with partial SU(3) symmetry was introduced in the 
framework of the symplectic shell model of nuclei~\cite{SymplM}.
The Hamiltonians were shown to be closely related to the deformation-inducing
quadrupole-quadrupole interaction and to possess both mixed-symmetry and 
solvable pure-SU(3) rotational bands.
For the example of the (prolate) deformed light nucleus $^{20}$Ne, it was
demonstrated that various features of the quadrupole-quadrupole interaction
can be reproduced with a particular parametrization of the PDS Hamiltonians.
In that work, the partial dynamical symmetry was identified directly
at the fermion level.
It is also possible to start with a bosonic PDS Hamiltonian and map the bosonic
generators into fermionic generators of the same algebra.
This approach was taken by Mamistvalov~\cite{Manana99}, who studied partial
symmetry in a schematic SU(2)$\times$SU(2)-type Lipkin model. 
Very recently, partially solvable shell-model Hamiltonians with 
seniority-conserving interactions were investigated by Rowe and 
Rosensteel~\cite{RoweRosen01}.

It is the purpose of this work to investigate the fermionic PDS Hamiltonians
presented in Ref.~\cite{Escher00PRL} in more detail.
Specifically, the construction process for the pure eigenstates is outlined 
and analytic expressions for the energies of pure states and the strengths of 
E2 transitions between these states are given.
Properties of the special solvable states are discussed and an application to 
the oblate deformed light nucleus $^{12}$C and the prolate nucleus $^{20}$Ne 
are presented.
Moreover, an application to $^{24}$Mg demonstrates the relevance of the 
PDS concept for well-deformed, triaxial nuclei.
 
In the next section, the symplectic shell model (SSM) is reviewed.
In Section~\ref{Sec:PDSHamiltonian}, a family of symplectic Hamiltonians with 
partial SU(3) symmetry is introduced and their relation to the 
quadrupole-quadrupole interaction is established. 
Properties of the special eigenstates of the PDS Hamiltonians are discussed 
in Section~\ref{Sec:States}, and applications to realistic nuclear systems 
are presented in Section~\ref{Sec:Applications}.
In Section~\ref{Sec:IBMComp}, the fermionic PDS Hamiltonians are compared
to the earlier introduced bosonic PDS 
Hamiltonians~\cite{Leviatan96a,Sinai99a,Sinai99b}, 
and Section~\ref{Sec:Summary} summarizes our work.  
Appendix A contains further relevant material regarding SU(3) coupling 
coefficients and reduced matrix elements and Appendix B presents expressions
for matrix elements of operators employed in the calculations.

\section{The Symplectic Shell Model}
\label{Sec:SSM}

The symplectic shell model (SSM) is an algebraic, fermionic, shell-model
scheme which includes multiple $2 \hbo$ one particle-one hole excitations.
It includes all essential observables for a description of nuclear
monopole and quadrupole collective vibrations as well as for rigid and 
irrotational flow rotations.
Since the model allows for intershell excitations and since its observables 
are expressible in microscopic shell-model terms, it provides a multi-shell 
realization of the nuclear shell model~\cite{SymplM}.

\subsection{Symplectic Generators}
\label{Subsec:SGen}

The symmetry algebra of the symplectic scheme is spanned by one-body operators 
which are bilinear products in the (relative) position ($x_{si}$, $i=1,2,3$, 
$s=1, \ldots, A-1$) and momentum ($p_{si}$) observables:
\bea
Q_{ij} &=& \sum_s x_{si} x_{sj}  \nonumber \\
K_{ij} &=& \sum_s p_{si} p_{sj}  \\
L_{ij} &=& \sum_s (x_{si} p_{sj} - x_{sj} p_{si})  \nonumber \\
S_{ij} &=& \sum_s (x_{si} p_{sj} + p_{si} x_{sj})  \;\; ,  \nonumber 
\eea
where $A-1$ is the number of Jacobi `particles' remaining after removal of the
center-of-mass contribution.
Together the operators generate the 21-dimensional symplectic algebra $sp$(6,R), that is, 
the Lie algebra of linear transformations which preserve a skew-symmetric bilinear 
form on a six-dimensional real vector space.
It is the smallest Lie algebra that  contains both the quadrupole moments and the 
many-nucleon kinetic energy, and it has several physically relevant subalgebras.
These include $gcm$(3) and its subalgebra $[R^5]so$(3), associated with the Geometric 
Collective Model and its rotational limit, respectively, the algebra $gl$(3,R) of 
the general linear motion group, as well as $su$(3) and its subalgebra $so$(3), 
associated with the Elliott model and the rotation group, respectively.
The $sp$(6,R) algebra furthermore includes the canonical subalgebras $sp$(2,R) and $sp$(4,R), 
which have been studied by Arickx et al.~\cite{Arickx82}, and by Peterson and Hecht 
\cite{Peter80}, respectively, as possible approximations to the full three-dimensional
symplectic model.

For many purposes, it is advantageous to express the symplectic generators in terms
of harmonic oscillator boson creation and annihilation operators 
$b_{si}^{\dagger}=(x_{si}-ip_{si})/\sqrt{2}$ and $b_s=(x_{si}+ip_{si})/\sqrt{2}$.
The symplectic generators may then be expressed as one-body operators which 
are quadratic in the oscillator bosons~\cite{Rosen80b}:
\bea
A_{ij} &=& \frac{1}{2} \sum_s b^{\dagger}_{si} b^{\dagger}_{sj} \nonumber\\
B_{ij} &=& \frac{1}{2} \sum_s b_{si} b_{sj}  \label{SymGenCart}\\
C_{ij} &=& \frac{1}{2} \sum_s ( b^{\dagger}_{si} b_{sj}+ b_{sj} b^{\dagger}_{si} )
\nonumber \; .
\eea
Alternatively, one may use the spherial components of the oscilator bosons, 
$b^{\dagger(10)}_{s, 1, \pm 1} = \mp \frac{1}{\sqrt{2}} 
( b_{s1}^{\dagger} \pm i b_{s2}^{\dagger})$,
$b^{\dagger(10)}_{s, 1, 0} = b_{s3}^{\dagger}$, and 
$\tilde{b}^{(01)}_{s, 1, \pm 1} = \mp \frac{1}{\sqrt{2}} ( b_{s1} \pm i b_{s2})$, 
$\tilde{b}^{(01)}_{s, 1, 0} = b_{s3}$, to write the generators as SU(3) tensor 
operators~\cite{Escher98b,RosenProc91}:
\bea
\begin{array}{lclrc}
\hat{H}_{0} &=& \sqrt{3} \sum_s \{ b_s^{\dagger (10)} \times
\tilde{b}_s^{(01)} \}^{(00)}_{00} + \frac{3}{2}(A-1) & & 
\\ \\  
\hat{C}^{(11)}_{lm} &=& \sqrt{2} \sum_s \{ b_s^{\dagger (10)} \times
\tilde{b}_s^{(01)} \}^{(11)}_{lm} & &  (l=1,2)
\\ \\
\hat{A}^{(20)}_{lm} &=& \frac{1}{\sqrt{2}} \sum_s \{ b_s^{\dagger (10)}
\times b_s^{\dagger (10)} \}^{(20)}_{lm} & & (l=0,2)
\\ \\
\hat{B}^{(02)}_{lm} &=& \frac{1}{\sqrt{2}} \sum_s \{ \tilde{b}_s^{(01)}
\times \tilde{b}_s^{(01)} \}^{(02)}_{lm} & & (l=0,2)  \; \; .
\end{array}
\eea
The notation $T^{\lm}_{lm}$ indicates that the operator $T$ possesses good SU(3) 
[superscript $\lm$] and SO(3) [subscript $lm$] tensorial properties.
Since $b_s^{\dagger}b_s^{\dagger}$ adds two quanta to particle $s$, thereby
moving it up across two major oscillator shells, $\hat{A}^{(20)}_{lm}$ creates a
$2 \hbo$ excitation in the system.
Analogously, $\hat{B}^{(02)}_{lm}$, which is related to $\hat{A}^{(20)}_{lm}$
by Hermitean conjugation, 
$\hat{B}^{(02)}_{lm} = (-1)^{l-m} (\hat{A}^{(20)}_{l-m})^{\dagger}$, 
annihilates a $2 \hbo$ excitation. 
The $\hat{C}^{(11)}_{lm}$ act only {\em within} a major harmonic oscillator shell.
They generate the group SU(3) of the well-known Elliott model~\cite{Elliott58}:
\bea
\sqrt{3} \hat{C}^{(11)}_{2m} &=& Q^E_{2m} \equiv \sqrt{\frac{4\pi}{5}} 
\sum_s (r^2_s Y_{2m} (\hat{r}_s) + p^2_s Y_{2m} (\hat{p}_s) ) \;\;\; (m=0,\pm1,\pm2) \; , 
\nn \\ \label{eq:SU3Gen} \\
\hat{C}^{(11)}_{1q}&=&\hat{L}_q \;\;\; (q=0,\pm1) \; , \nn
\eea
where $Q^E_{2m}$ denotes the symmetrized quadrupole operator of Elliott, which 
does not couple different major shells, and $\hat{L}_q$ is the orbital angular 
momentum operator.
The harmonic oscillator Hamiltonian, $\hat{H}_0 = \sum_{i=1}^{3} \hat{C}_{ii}$, 
is a SU(3) scalar and generates U(1) in U(3) $=$  SU(3) $\times$ U(1).

Alternatively, one can realize the symplectic generators in terms of fermionic
creation and annihilation operators~\cite{Escher98b}:
\bea
\begin{array}{lclrc}
\hat{C}^{(11)}_{lm} &=&  
\sum_{\eta} \sqrt{\frac{1}{6} \eta (\eta+1) (\eta+2) (\eta+3)} 
\{ a_{\eta}^{\dagger} \times \tilde{a}_{\eta} \}^{(11)S=0}_{lm\;\;\Sigma=0} 
 +  \hat{{\cal O}}_{C}^{cm}(A) & & 
\\ \\
\hat{A}^{(20)}_{lm} &=& 
\sum_{\eta} \sqrt{\frac{1}{12} (\eta+1) (\eta+2) (\eta+3) (\eta+4)} 
\{ a_{\eta+2}^{\dagger} \times \tilde{a}_{\eta} \}^{(20)S=0}_{lm\;\;\Sigma=0}
& + & \hat{{\cal O}}_{A}^{cm}(A)                                        
\\ \\
\hat{B}^{(02)}_{lm} &=& 
\sum_{\eta} \sqrt{\frac{1}{12} (\eta+1) (\eta+2) (\eta+3) (\eta+4)} 
\{ a_{\eta}^{\dagger} \times \tilde{a}_{\eta+2} \}^{(02)S=0}_{lm\;\;\Sigma=0} 
& + & \hat{{\cal O}}_{B}^{cm}(A)
\end{array}
\eea
where $a_{\eta l m 1/2 \sigma}^{\dagger}$  
($\tilde{a}_{\eta l m 1/2 \sigma} = (-1)^{\eta+l+m+1/2+\sigma}
a_{\eta l -m 1/2 -\sigma}$) 
is a single-particle creation (annihilation) operator, which produces (destroys) 
a fermion with angular momentum $l$, projection $m$ and spin $1/2$, projection 
$\sigma$ in the $\eta$-th major oscillator shell.
The sums run over all shells, and the coupling to total spin $S=0$ with projection 
$\Sigma=0$ reflects the fact that the generators act on spatial degrees of freedom 
only.
The operators $\hat{{\cal O}}^{cm}(A)$ remove the spurios center-of-mass content 
from the generators.  Details regarding the fermionic realization of Sp(6,R) can
be found in Ref.~\cite{Escher98b}.

\subsection{Symplectic Basis States}
\label{Subsec:SBas}

A basis for the symplectic model is generated by applying symmetrically
coupled products of the 2$\hbo$ raising operator $\hat{A}^{(20)}$ with
itself to the usual $0 \hbo$ many-particle shell-model states.
Each $0 \hbo$ starting configuration is characterized by the distribution of 
oscillator quanta into the three cartesian directions, $\{ \s_1,\s_2,\s_3 \}$,
where $\s_1 \geq \s_2 \geq \s_3$. 
Here $\sigma_i$ denotes the eigenvalue of the U(3) weight operator
$C_{ii}=\sum_s (b^{\dagger}_{si} b_{si} + 1/2)$, which essentially counts the 
number of oscillator bosons in the $i$-th direction of the system.
Since $s=1,2,\ldots,A-1$, it follows that the $\sigma_i$ are half-integer numbers 
for even-A and integers for odd-A nuclei.
Equivalently, one may employ quantum numbers $N_{\s} \lms$, where
$\la_{\s} = \s_1 - \s_2$, $\mu_{\s} = \s_2 - \s_3$ are the Elliott SU(3) labels,
and $N_{\s} = \s_1 +\s_2 +\s_3$ is the eigenvalue of the harmonic oscillator 
Hamiltonian, $\hat{H}_0=\hat{C}_{11}+\hat{C}_{22}+\hat{C}_{33}$, which takes the 
minimum value consistent with the Pauli Exclusion Principle. 
Each such set of U(3) quantum numbers uniquely determines an irreducible 
representation (irrep) of the symplectic group, since it characterizes a Sp(6,R) 
lowest weight state. 
Any component of the symplectic lowering operator $B^{(20)}$ (and of $\hat{C}_{ij}$ 
with $i < j$) annihilates such a lowest weight state.

\begin{figure}[htp]
\vskip 3cm
\hskip 1cm
\hbox{
\epsfxsize=4.5 true in
\epsfbox{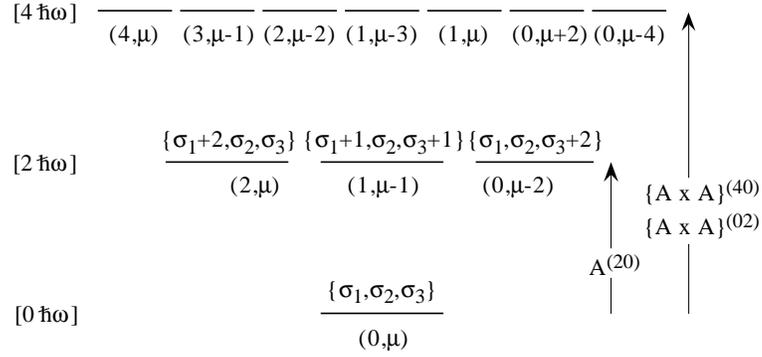}
}
\caption{
Basis construction in the symplectic model.
SU(3)-coupled products of the raising operator $\hat{A}^{(20)}$ with
itself act on an Elliott starting state with $\lms = (0,\mu)$
($\{\s_1,\s_2 =\s_1,\s_3\}$) to generate symplectic $2\hbo$, $4\hbo$,
$\ldots$ excitations.
Also shown are the SU(3) labels $\lm$ and quanta distributions
$\{\om_1,\om_2,\om_3\}$ for some excited states.
}
\label{SymplIrrep}
\end{figure}

In contrast, application of the symplectic generator  $\hat{A}^{(20)}$ allows 
one to successively build a basis for the Sp(6,R) irrep under consideration: 
The product of $N/2$, $N=0,2,4,\ldots$, raising operators $\hat{A}^{(20)}$ is
multiplicity-free and generates $N\hbo$
excitations for each starting configuration $N_{\s} \lms$.
Each such product operator ${\cal P}^{N (\la_n,\mu_n)}$ can be labeled
according to its U(3) content, $\{n_1,n_2,n_3\}$ or $N (\la_n,\mu_n)$, where 
$(\la_n , \mu_n)$ ranges over the set 
\bea
\Omega &=& \{ (n_1-n_2,n_2-n_3) |
n_1 \geq n_2 \geq n_3 \geq 0; N=n_1+n_2+n_3; n_1,n_2,n_3 \: 
\mbox{even integers} \} \, .
\eea
The raising polynomial ${\cal P}^{N (\la_n,\mu_n)}$ is then coupled with
$| N_{\s} \lms \rangle$ to good SU(3) symmetry $\rho \lmw$, with $\rho$
denoting the multiplicity in the coupling $(\la_n,\mu_n) \otimes \lms$.
The quanta distribution in the resulting state is given by
$\{ \om_1,\om_2,\om_3 \} $, with 
$N_{\om} \equiv N_{\s} + N = \om_1 + \om_2 + \om_3$,
$\om_1 \geq \om_2 \geq \om_3$, and $\la_{\om} = \om_1 - \om_2$,
$\mu_{\om} = \om_2 - \om_3$.
The states of the Sp(6,R) $\supset$ SU(3) basis are thus labeled by three 
types of U(3) quantum numbers: 
$\Gamma_{\s} \equiv$ $\{ \s_1,\s_2,\s_3 \} =$ $N_{\sigma} \lms$, 
the symplectic bandhead or Sp(6,R) lowest weight U(3) symmetry, which specifies 
the Sp(6,R) irreducible representation; 
$\Gamma_n \equiv$ $\{n_1,n_2,n_3\} =$ $N(\la_n \mu_n)$, the U(3) symmetry of 
the raising polynomial; 
and $\Gamma_{\omega} \equiv$ $\{\om_1,\om_2,\om_3\} =$ $N_{\omega} \lmw$, 
the U(3) symmetry of the coupled product.
A given symplectic representation space $N_{\sigma} \lms$ is infinite
dimensional, since $N/2$, the number of oscillator excitations, can take any 
non-negative integer value.  In practical applications, one must therefore either
truncate the symplectic Hilbert space, or restrict oneself to interactions
and observables for which the matrix elements depend solely on the symplectic 
irrep and can be calculated analytically.
The basis state construction is schematically illustrated in
Fig.~\ref{SymplIrrep} for a typical Elliott starting state with
$(\lambda_{\s},\mu_{\s}) = ( 0,\mu )$.
A similar figure for $(\lambda_{\s},\mu_{\s}) = (\la ,0 )$ is given in
Ref.~\cite{Escher00PRL}.

To complete the basis state labeling, additional quantum numbers $\alpha$
are required.
This can be accomplished by reducing Sp(6,R) $\supset$ SU(3) states with
respect to the subgroup U(1) $\times$ SU(2) of SU(3) and assigning labels
$\alpha = \varepsilon \Lambda M_{\Lambda}$
\footnote{
Here $\varepsilon$, the eigenvalue of $Q^E_{20}$, gives the U(1) content and 
the SU(2) irrep is characterized by $\Lambda$ with projection $M_{\Lambda}$.
}.
This SU(2) subgroup, however, is not the physical orbital angular momentum
subgroup SO(3) of SU(3).  States with good angular momentum values can be 
obtained from the SU(3) $\supset$ U(1) $\times$ SU(2) (canonical) basis by 
projection~\cite{Elliott58,JPD73a}.
The associated quantum numbers are $\alpha = \kappa L M$, where $\kappa$ is 
a multiplicity index, which enumerates multiple occurrences of a particular 
$L$ value in the SU(3) irrep $\lm$ from 1 to $\kappa^{max}_L \lm$,
\bea
\kappa^{max}_L \lm &=& [(\la+\mu+2-L)/2] - [(\la+1-L)/2] - [(\mu+1-L)/2]  \; ,
\label{eq:KapMax}
\eea
where [$\ldots$] is the greatest non-negative integer function~\cite{Lopez90}.
The $\kappa^{max}_L \lm$ occurrences of $L$ can be distinguished in a variety
of ways.
The physically most significant scheme is that of Elliott~\cite{Elliott58},
in which case the projection of $L$ along the body-fixed 3-axis, denoted K,
is used to sort the $L$-values into the familiar K-bands of the rotational
model.
Unfortunately, states defined in this manner are not orthonormal with respect
to the multiplicity quantum number K. 
To avoid the resulting complications, such as working with non-hermitean matrices,
the Elliott basis is usually orthonormalized using a Gram-Schmidt process.
Vergados~\cite{Vergados}, for example, gives a prescription to construct 
orthogonal basis states in a systematic manner for all $\lm$,
such that the physical interpretation of K as a band label can be approximately
maintained
\footnote{
Vergados projects from a state with $\varepsilon=\varepsilon_{min}=-\la-2\mu$, 
$\Lambda = \la/2$, $M_{\Lambda} = \la/2$ for $\la \geq \mu$ and 
$\varepsilon=\varepsilon_{max}=2\la+\mu$, $\Lambda=\mu/2$, $M_{\Lambda}= -\mu/2$ 
for $\la < \mu$ and employs the `Elliott rule' to determine the possible K values, 
K = min$\lm$, min$\lm$ - 2, $\ldots$ , 1 or 0, and angular momenta, 
$L$ = K, K+1, K+2, $\ldots$, K+max$\lm$ for K $\neq$ 0 and 
$L$=max$\lm$, max$\lm$-2, $\ldots$, 1 or 0 for K=0.
It is also possible to project from $\varepsilon=\varepsilon_{max}$, 
$\Lambda=\mu/2$, and $M_{\Lambda}= +\mu/2$ or $\varepsilon=\varepsilon_{min}$,
$\Lambda=\la/2$, $M_{\Lambda}=-\la/2$.
Draayer {\em et al.}~\cite{JPD73a,JPDSU3Basis} discuss the different projection
possibilities and give rules analogous to the Elliott rule for determining the 
K and $L$ content of a given SU(3) irrep $\lm$. 
}.
In the present work, we employ the orthonormal basis of Vergados.
For simplicity, however, we use the running index 
$\kappa = 1, 2, \ldots, \kappa^{max}_L$
to distinguish multiple occurrences of $L$ in a given SU(3) irrep $\lm$
and list the corresponding Vergados labels where appropriate.
The dynamical symmetry chain and the associated quantum labels of the above
scheme are given by~\cite{SymplM}:
\bea
\begin{array}{ccccccc}
    Sp(6,R) &\supset& U(3) &\supset& SO(3) &\supset& SO(2) \\
 \\
    N_{\sigma}(\lambda_{\sigma},\mu_{\sigma}) & N(\lambda_n,\mu_n) \rho &
    N_{\omega}(\lambda_{\omega},\mu_{\omega})& \kappa & L && M
\end{array}
\label{eq:DSBasis}
\eea                                           
When applying the formalism to realistic nuclei, we assign rotational band
labels according to the calculated B(E2) rates.

The quadratic Casimir operators of SU(3) and Sp(6,R),
\bea
\hat{C}_{SU3} &=& \frac{1}{2}\left [ C^{(11)}_2 \cdot C^{(11)}_2 
                 + C^{(11)}_1 \cdot C^{(11)}_1 \right]
\label{eq:SU3Cas} \\ 
\hat{C}_{Sp6} &=& - 2\hat{A}^{(20)}_0\hat{B}^{(02)}_0 
                  - 2\hat{A}^{(20)}_2 \cdot \hat{B}^{(02)}_2 
                  + \hat{C}_{SU3} +
                  \frac{1}{3}\hat{H_0}^{2} - 4\hat{H_0}  \; ,
\label{eq:Sp6Cas}
\eea
have the following eigenvalues in the dynamical symmetry basis:
\bea
\langle \hat{C}_{SU3} \rangle [\lm]        
  &=& 2 (\la^2 + \mu^2 + \la \mu + 3\la + 3\mu)/3 
\label{eq:SU3CasVal}  \\ \nn \\
\langle \hat{C}_{Sp6} \rangle [N_{\s}\lms] 
  &=& 2(\la_{\s}^2+\mu_{\s}^2+\la_{\s}\mu_{\s}+3\la_{\s}+3\mu_{\s})/3
           +N_{\s}^2/3-4N_{\s}  \; .
\label{eq:Sp6CasVal}
\eea

\begin{figure}[hbtp]
\begin{center}
\epsfxsize=5 true in
\epsfbox{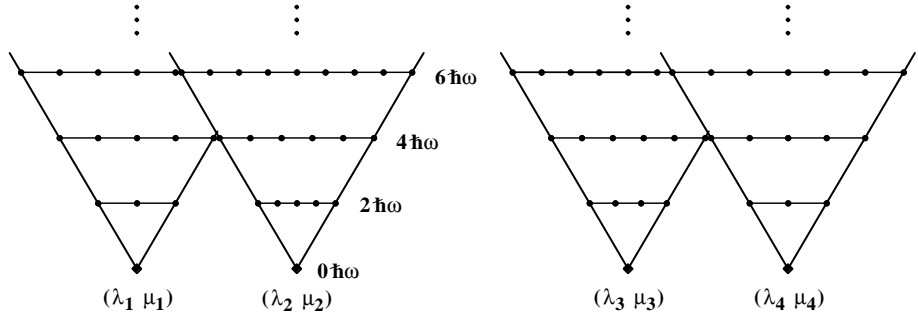}
\end{center}
\vskip -10 pt
\caption{
Symplectic shell model space.
The schematic plot illustrates a model space with multiple symplectic
representations.
Each `cone' corresponds to a Sp(6,R) irrep and is uniquely characterized
by U(3) quantum numbers $N_{\s}\lms$, where $\lms$ denotes the Elliott SU(3)
quantum labels for the associated 0$\hbo$ shell model configuration.
For a given starting representation $\lms$ ($\s = 1,2,3,4$ here), one obtains multiple SU(3)
configurations, $\lmw$, at each $N\hbo$ level of excitation ($N>0$),
indicated here by small filled circles.
}
\label{Sp6Cones}
\end{figure}

The collection of all $0 \hbo$ configurations provides a complete Hilbert space
for the Elliott SU(3) submodel of the SSM and is referred to as the $0 \hbo$ 
horizontal shell-model space.
The set of states built on a given U(3) irrep $N_{\sigma} (\la_{\sigma} \mu_{\sigma})$ 
is called the vertical extension of that irrep.
Each vertical extension can be partitioned into horizontal slices with the 
states within the $N/2$-th slice representable as a homogeneous polynomial of 
degree $N/2$ in the $\hat{A}^{(20)}$ tensors acting on the parent $0 \hbo$ 
configuration (see also Fig.~\ref{Sp6Cones}).
Interactions can thus be classified according to their effect on this structure; 
pairing, for example, causes horizontal mixing, both within each `cone' (symplectic 
irrep) and between different cones, while the quadrupole-quadrupole interaction 
induces horizontal and vertical mixing, but does not connect different cones. 

\subsection{Symplectic Hamiltonians}
\label{Subsec:SHam}

A primary goal of the symplectic shell model is to achieve a microscopic description 
of deformed nuclei.
These nuclei exhibit collective behavior, that is, modes of excitation 
in which an appreciable fraction of the nucleons in the system 
participate in a coherent manner, as, for example, is the case for 
rotations.
An appropriate Hamiltonian for describing rotational phenomena within 
the symplectic model consists of the harmonic oscillator, which provides 
the background shell structure, the quadrupole-quadrupole interaction, 
$Q_2 \cdot Q_2$, and a residual interaction that should include, for
example, single-particle spin-orbit and orbit-orbit terms, as well as 
pairing and other interactions.
However, most applications of the theory are much less ambitions than 
this, restricting the interaction to terms that can be expressed solely 
in terms of generators of the symplectic 
algebra~\cite{SymplM,JPD92b,Rosen84,Casta89a}. 
Interactions of the latter form do not mix different symplectic irreps 
and therefore the Hamiltonian matrix for such interactions becomes 
block-diagonal.
Indeed, in most practical applications the Hilbert space of the 
system is truncated to one single symplectic representation.
This is accomplished by selecting the vertical slice (symplectic irrep) 
constructed from the leading starting irrep of the $0 \hbo$ space.
The leading irrep is defined to be the U(3) representation, $N_{\sigma} \lms$, 
from the lowest layer with the most symmetric spatial permutation symmetry 
consistent with the Pauli principle, and the maximal possible SU(3) Casimir 
operator value, $\langle \hat{C}_{SU3} \rangle [\lms]$.
For $^{12}$C, for instance, the leading irrep is given by
$N_{\s}$ $(\lambda_{\s},\mu_{\s})$ = 24.5 (0,4), which corresponds to the
symplectic weights $\sigma_1$=$\sigma_2$=9.5, $\sigma_3$=5.5; 
for $^{20}$Ne, one finds $N_{\s}$ $(\lambda_{\s},\mu_{\s})$ = 48.5 (8,0), 
since $\sigma_1$=21.5, $\sigma_2$=$\sigma_3$=13.5~\cite{SymplM}; 
and $^{24}$Mg has $N_{\s}$ $(\lambda_{\s},\mu_{\s})$ = 62.5 (8,4), that is, 
$\sigma_1$=27.5 $\sigma_2$=19.5, $\sigma_3$=15.5~\cite{Rosen84}.      
The single-symplectic irrep approximation is a sensible choice for nuclear 
systems which have a dominant quadrupole-quadrupole force, since this 
interaction does not mix symplectic representations and favors states with 
large $\langle \hat{C}_{SU3} \rangle [\lms]$ values.

A typical Hamiltonian for a calculation in a space truncated in the
manner described above,  is given by a harmonic oscillator term,
$H_0$, plus a collective potential, and a residual interaction:
\bea
H &=& H_0 + V_{coll} + V_{res}  \; .
\eea
We choose the collective potential to be a simple quadratic, rotationally
invariant, function of the microscopic quadrupole moment\footnote{
Higher order rotational scalars in $Q_2$ can be included in $V_{coll}$
in order to accomodate more complicated potential forms, e.g.\
a cubic term introduces a $\gamma$-dependence into the potential.
},
$ Q_{2m}=\sqrt{\frac{16\pi}{5}} \sum_s r^2_s Y_{2m} (\hat{r}_s)$, namely
\bea
V_{coll} &=& - \chi Q_2 \cdot Q_2  \; .
\eea
The quadrupole-quadrupole interaction is a standard ingredient in models 
that aim at reproducing rotational spectra and nuclear deformations.  
It emerges (apart from a constant) as a leading contribution in the 
multipole expansion of a general two-body force.  
It mixes states from different oscillator shells, since the quadrupole operator 
has non-vanishing matrix elements between shells differing by zero or two 
oscillator quanta.
A major strength of the symplectic model is its ability to fully accommodate
the action of the quadrupole operator, which can be written in terms of
symplectic generators:
\bea
Q_{2m} &=& 
\sqrt{3} ( \hat{C}^{(11)}_{2m} + \hat{A}^{(20)}_{2m} + \hat{B}^{(02)}_{2m} ) 
\; . 
\eea
As a result, the model is able to reproduce intra-band and inter-band E2
transition strengths between low-lying, as well as giant resonance, states
without introducing proton and neutron effective charges.

The effective residual interaction, $V_{res}$, is included to replace 
non-collective components of a more realistic Hamiltonian and the neglected 
effects of couplings to other Sp(6,R) representations.
As in previous works, we choose $V_{res}$ to be a rotationally invariant
function of the SU(3) generators.
For prolate and oblate nuclei we use:
\bea
V_{res} &=& d_2 \hat{L}^2 + d_4 \hat{L}^4 \; ,
\eea  
where $\hat{L}$ denotes the angular momentum operator, Eq.~(\ref{eq:SU3Gen}).
This allows us to reproduce the energy splittings between states of a rotational 
band.
For triaxial nuclei, such as $^{24}$Mg, it becomes necessary to include further
terms, in order to reproduce the experimentally observed `K-band splitting', 
the energy differences found between states with the same angular momentum 
but different K-band assignments.
This can be achieved by including `SU(3) $\supset$ SO(3) integrity basis' 
operators $\hat{X}_3 \equiv (\hat{L} \times Q^E)_{(1)} \cdot \hat{L}$ and 
$\hat{X}_4 \equiv (\hat{L} \times Q^E)_{(1)} \cdot (\hat{L} \times Q^E)_{(1)}$ 
in the residual interaction~\cite{JPD85}: 
\bea
V_{res}' &=& c_3 \hat{X}_3 + c_4 \hat{X}_4 + d_2 \hat{L}^2 + d_4 \hat{L}^4 \; .
\eea

The evaluation procedure for the matrix elements of the symplectic generators 
$A^{(20)}$, $B^{(02)}$, and $C^{(11)}$, and of the integrity basis operators 
$\hat{X}_3$ and $\hat{X}_4$ is discussed in Appendix B.

\section{PDS Hamiltonians and the quadrupole-quadrupole interaction}
\label{Sec:PDSHamiltonian}                                

In this section we introduce a family of fermionic Hamiltonians with
partial dynamical symmetry.
Motivated by the fact that a realistic quadrupole-quadrupole interaction
breaks SU(3) symmetry within a given major oscillator shell, we define
a family of Hamiltonians, $H(\beta_0,\beta_2)$, which allows us to 
study the features of the symmetry-breaking terms in some detail.
The new Hamiltonians do not couple different oscillator shells and, for a
particular choice of the parameters $\beta_0$ and $\beta_2$,
reduce to a form which is closely related to the quadrupole-quadrupole
interaction restricted to a shell.
We prove that this family of Hamiltonians exhibits partial SU(3) symmetry
and give rules for determining the `special' irreps and the associated
pure eigenstates.

In the symplectic shell model, the quadrupole-quadrupole interaction can be 
expressed in terms of symplectic generators~\cite{Rosen90b}:
\bea
Q_2 \cdot Q_2 &=& 3(\hat{C}_2 + \hat{A}_2 + \hat{B}_2 )
                  \cdot (\hat{C}_2 + \hat{A}_2 + \hat{B}_2) \; .
\eea
Employing the commutation relations 
$\hat{B}_{2}\cdot \hat{A}_{2} - \hat{A}_{2}\cdot \hat{B}_{2} = 
\frac{10}{3}\hat{H}_0$ and 
$\hat{B}_{2}\cdot \hat{C}_{2} - \hat{C}_{2}\cdot B_{2} = 
\frac{20}{\sqrt{6}} \hat{B}_{0}$, 
given in Ref.~\cite{Rosen90b}, this can be rewritten as:
\bea
Q_2 \cdot Q_2 &=& 3 \hat{C}_2 \cdot \hat{C}_2 + 6 \hat{A}_{2}\cdot \hat{B}_{2} 
                 + 10\hat{H_0}   \nonumber \\
&&  + \left[ \left( 6 \hat{C}_2 \cdot \hat{B}_2 + 10\sqrt{6}\hat{B}_0 
    + 3\hat{B}_{2}\cdot \hat{B}_{2} \right) + \mbox{H.c.} \right]
\label{Eq:QQGeorge}
\eea
where $3 \hat{C}_2 \cdot \hat{C}_2 = Q_2^E \cdot Q_2^E$ and H.c.\/ denotes 
the Hermitian conjugate of the expression in parentheses.
The first three terms in the expansion, Eq.~(\ref{Eq:QQGeorge}), act solely 
within a major harmonic oscillator shell, while the second line connects 
states differing in energy by $\pm 2\hbo$ and $\pm 4\hbo$.
It is primarily the presence of the multi-$\hbo$ correlations that
differentiates the SSM from the Elliott SU(3) model.
The symplectic model allows for coherent multi-shell admixtures in its wave 
functions and thus achieves the experimentally observed nuclear deformation 
and absolute B(E2) rates. 
In contrast, the Elliott model requires effective charges, since it employs
the algebraic (or Elliott) quadrupole-quadrupole interaction,
\bea
Q_2^E \cdot Q_2^E &=& 6 \hat{C}_{SU3} - 3 \hat{L}^2  \; ,
\eea
which does not connect different oscillator shells.

Although matrix elements of $Q_2$ and $Q^E_2$ are identical within a harmonic 
oscillator shell, the corresponding quadrupole-quadrupole interactions exhibit 
differences here as well: The $\hat{C}_2 \cdot \hat{C}_2$ and $\hat{H_0}$ terms 
in the expansion, Eq.~(\ref{Eq:QQGeorge}), are diagonal in the dynamical symmetry 
basis, Eq.~(\ref{eq:DSBasis}), whereas $A_{2}\cdot B_{2}$ contains contributions 
which mix different SU(3) irreps.
This follows from the relations:
\bea
\hat{A}_0 \hat{B}_0 &\equiv& \hat{A}^{(20)}_0 \hat{B}^{(02)}_0
= \frac{1}{\sqrt{6}} \{ \hat{A} \times \hat{B} \}_0^{(00)}
- \sqrt{\frac{5}{6}} \{ \hat{A} \times \hat{B} \}_0^{(22)}   \;\; ,
\nn \\
\hat{A}_2 \cdot \hat{B}_2 &\equiv& \hat{A}^{(20)}_2 \cdot \hat{B}^{(02)}_2
= \frac{5}{\sqrt{6}} \{ \hat{A} \times \hat{B} \}_0^{(00)}
+ \sqrt{\frac{5}{6}} \{ \hat{A} \times \hat{B} \}_0^{(22)}   \;\; ,
\label{eq:ABRecoupl}
\eea
where
\bea
\{ \hat{A} \times \hat{B} \}_0^{(00)} &=& \frac{1}{2\sqrt{6}}
( \hat{C}_{SU3} + \frac{1}{3} \hat{H}_0^2
- 4 \hat{H}_0 - \hat{C}_{Sp6} ) \; .
\eea                                          
The term $\{ \hat{A} \times \hat{B} \}_0^{(00)}$ is a SU(3) scalar, but 
$\{ \hat{A} \times \hat{B} \}_0^{(22)}$ breaks SU(3) symmetry.
Within a major oscillator shell, it is mainly this symmetry-breaking term that 
distinguishes the action of $Q_2 \cdot Q_2$ from the effect of the Elliott 
interaction, $Q_2^E \cdot Q_2^E$, which respects the symmetry.

To explore this latter aspect in more detail, we rewrite the collective 
quadrupole-quadrupole interaction as follows:
\bea
Q_2 \cdot Q_2 &=& 9 \hat{C}_{SU3} - 3 \hat{C}_{Sp6} +
\hat{H}_0^2 - 2 \hat{H}_0 - 3 \hat{L}^2
- 6 \hat{A}_0 \hat{B}_0 \nonumber \\
&& + \{ \mbox{terms coupling different h.o.\/ shells} \} \; . 
\label{Eq:QQ}
\eea
The quadratic Casimir invariants of SU(3), $\hat{C}_{SU3}$, and of Sp(6,R),
$\hat{C}_{Sp6}$, and their eigenvalues, are given in 
Eqs.~(\ref{eq:SU3Cas})--(\ref{eq:Sp6CasVal}).
In order to focus on the action of $Q_2 \cdot Q_2$ within a harmonic oscillator
shell, we introduce the following family of rotationally invariant Hamiltonians:
\bea
\lefteqn{H(\beta_0,\beta_2) = \beta_0 \hat{A}_0 \hat{B}_0
+ \beta_2 \hat{A}_2 \cdot \hat{B}_2 }
\label{Eq:Hpds} \nn \\
&& = \frac{\beta_2}{18} ( 9\hat{C}_{SU3} - 9\hat{C}_{Sp6}
+ 3\hat{H}_0^2 - 36\hat{H}_0 )
   + ( \beta_0 - \beta_2) \hat{A}_0 \hat{B}_0 \; .
\eea
For $\beta_0=\beta_2$, one recovers a Sp(6,R)$\supset$SU(3) dynamical symmetry 
Hamiltonian:  $H(\beta_0,\beta_2=\beta_0)$ contains only SU(3)-scalars, that is, 
it does not mix different SU(3) irreps.  
Furthermore, all eigenstates at a given $N\hbar\omega$ excitation which belong
to the same symplectic and SU(3) irreps are degenerate.
Additional SO(3) rotational terms, such as $\hat{L}^2$ and $\hat{L}^4$ split
the degeneracies, but do not change the wave functions.  
For the special choice $\beta_0=12$, $\beta_2=18$, one finds that 
$H(\beta_0=12,\beta_2=18)$ is closely related to the quadrupole-quadrupole
interaction: 
\bea
Q_2 \cdot Q_2 &=& H(\beta_0=12,\beta_2=18)+ const - 3 \hat{L}^2 \nn \\
&& + \{ \mbox{terms coupling different h.o.\/ shells} \}  \;\; ,
\eea
where the value of $const = 6\hat{C}_{Sp6} - 2\hat{H}_0^2 + 34\hat{H}_0$ is 
fixed for a given symplectic irrep $N_{\s}\lms$ and $N\hbo$ excitation.  
Although $H(\beta_0,\beta_2)$ does not couple different harmonic oscillator 
shells, it contains the SU(3)-symmetry breaking term 
$\{ \hat{A} \times \hat{B} \}_0^{(22)}$ and is therefore expected to exhibit 
in-shell behavior similar to that of $Q_2 \cdot Q_2$.

From Eq.~(\ref{eq:ABRecoupl}) it follows that $H(\beta_0,\beta_2)$ is generally
not SU(3) invariant. 
We will now show that $H(\beta_0,\beta_2)$ exhibits partial SU(3) symmetry.
Specifically, we claim that among the eigenstates of $H(\beta_0,\beta_2)$,
there exists a subset of solvable pure-SU(3) states, the SU(3)$\supset$SO(3) 
classification of which depends on both the Elliott labels 
$(\lambda_{\s},\mu_{\s})$ of the starting state and the symplectic excitation $N$.
In general, we find that all $L$-states in the starting configuration ($N=0$)
are solvable with good SU(3) symmetry $\lms$. 
For excited configurations, with $N>0$ ($N$ even), we distinguish two possible 
cases:
\begin{itemize}
\item[(a)] $\la_{\s} > \mu_{\s}$:
the pure states belong to $\lm = (\la_{\s}-N,\mu_{\s}+N)$ 
at the $N \hbar\omega$ level and have
$L = \mu_{\s}+N, \mu_{\s}+N+1, \ldots , \la_{\s}-N+1$
with $N=2,4, \ldots$ subject to $2N \leq (\la_{\s} - \mu_{\s} + 1)$.
\item[(b)] $\la_{\s} \leq \mu_{\s}$:
the special states belong to $\lm =(\la_{\s}+N,\mu_{\s})$
at the $N \hbar\omega$ level and have
$L = \la_{\s}+N, \la_{\s}+N+1, \ldots , \la_{\s}+N+\mu_{\s}$
with $N=2,4, \ldots$.
\end{itemize}

To prove the claim, it is sufficient to show that $\hat{B}_0$ annihilates 
the states in question (since $H(\beta_0=\beta_2)$ is diagonal in the 
dynamical symmetry basis).
For $N=0$ this follows immediately from the fact that the $0 \hbo$
starting configuration is a Sp(6,R) lowest weight which, by definition,
is annihilated by the lowering operators of the Sp(6,R) algebra.
The latter include the components of the generator $\hat{B}^{(02)}$. 

For $N>0$, we have to consider the action of $\hat{B}_0$ in more detail. 
Let $\{ \sigma_1,\sigma_2,\sigma_3 \}$ be the quanta distribution for a
0$\hbo$ state with $\la_{\s} > \mu_{\s}$.
An excited state with SU(3) character $\lm = (\la_{\s}-N,\mu_{\s}+N)$
must have the quanta distribution $\{ \s_1, \s_2+N, \s_3 \}$.
Acting with the rotational invariant $\hat{B}_0$ on such a state does not
affect the angular momentum, but removes two quanta from the 2-direction, 
giving a $(N-2) \hbo$ state with $(\la',\mu') = (\la_{\s}-N+2,\mu_{\s}+N-2)$.
Note that the symplectic generator $\hat{B}_0$ cannot remove quanta from the 
other two directions of this particular state, since this would yield a state 
which has fewer oscillator quanta in the 1- or 3-direction than the starting
(0$\hbo$) configuration, i.e. the resulting state would belong to a different 
symplectic irrep.
Comparing the number of occurrences of a given angular momentum value $L$ in 
$\lm$ at $N \hbo$ and $(\la',\mu')$ at $(N-2) \hbo$, one finds the following:
As long as $\la_{\s} - N + 1 \geq \mu_{\s}+N$ holds, the difference
$\Delta_L(N) \equiv \kappa_L^{max}\lm - \kappa_L^{max}(\la',\mu')$ is 1
for $L = \mu_{\s}+N, \mu_{\s}+N+1, \ldots, \la_{\s}-N+1$, and zero otherwise
(with $\kappa_L^{max}$ as defined in Eq.~(\ref{eq:KapMax})).
Therefore, when $\Delta_L(N)$=1, a linear combination $|\phi_L(N)\rangle =
\sum_{\kappa} c_{\kappa} | N\hbo (\la_{\s}-N,\mu_{\s}+N) \kappa L M \rangle$
exists such that $\hat{B}_0 |\phi_L(N)\rangle = 0$, and thus our claim
for family (a) holds.
 
The proof for family (b) can be carried out analogously.
Here the special irrep $\lm =(\la_{\s}+N,\mu_{\s})$ is obtained by adding
$N$ quanta to the 1-direction of the starting configuration.
In this case there is no restriction on $N$, hence family (b) is infinite.
Note that adding quanta to the 3-direction does {\em not} yield another
family of pure states, since the multiplicity for a given $L$-value in
the associated `special' irreps, $\lm = (\la_{\s},\m_{\s}-N)$, decreases 
as $N$ increases, i.e. $\Delta_L(N) \leq 0$ for all $L$ and $N$.

\section{Solvable states and their properties}
\label{Sec:States}
 
All $0\hbo$ states are eigenstates of $H(\beta_0,\beta_2)$.
They are unmixed and span the entire $\lms$ irrep. 
In contrast, for the excited levels ($N > 0$), the pure states span only 
part of the corresponding SU(3) irrep.
There are other states at each excited level which do not preserve the
SU(3) symmetry and therefore contain a mixture of SU(3) irreps.

To construct the pure states for $N>0$, we proceed as follows:
Let $\lm$ at $N\hbo$ be the irrep which contains a pure state with
angular momentum $L$ and projection $M$, $|\phi_{LM}(N)\rangle$.
This state can be written as:
\bea 
|\phi_{LM}(N)\rangle &=& \sum_{\kappa=1}^{\kappa^{max}_L \lm}
c_{\kappa}(L) | N\hbo \lm \kappa L M \rangle \; ,
\eea
where $\kappa^{max}_L \lm$ denotes the maximum multiplicity of $L$ in $\lm$,
Eq.~(\ref{eq:KapMax}).
Obviously, $|\phi_{LM}(N)\rangle$ is an unmixed eigenstate of $H(\beta_0,\beta_2)$ 
if $\langle \psi(N-2) | B_0 | \phi_{LM}(N)\rangle = 0$ holds for all states
$|\psi(N-2) \rangle$ at the $(N-2)\hbo$ level.
From the proof it follows that $B_0$ acting on states in the `special'
irrep $\lm$ at $N\hbo$ can only produce states belonging to the `special'
irrep $\lmp$ at $(N-2)\hbo$, hence 
$\langle (N-2)\hbo \lmp \kappa' L M | B_0 | \phi_L(N) \rangle = 0$ for
$\kappa'=1,2,\ldots,\kappa^{max}_L \lmp$ ensures that $|\phi_{LM}(N)\rangle$
is pure.
The $\kappa^{max}_L \lm$ coefficients $c_{\kappa}(L)$, which characterize the 
pure state, are thus uniquely determined by the $\kappa^{max}_L \lmp$ = 
$\kappa^{max}_L \lm - 1$ equations
\begin{equation}
\sum_{\kappa} c_{\kappa}(L) \langle (N-2)\hbo \lmp \kappa' L M | B_0 |
N\hbo \lm \kappa L M \rangle = 0
\label{Eq:CoeffDet1}
\end{equation}
and the normalization requirement $\sum_{\kappa} |c_{\kappa}(L)|^2 = 1$.
The proof given in the previous section guarantees the existence of a
solution.

Making use of the Wigner-Eckart theorem for SU(3) (see Appendix A),
the relations in Eq.~(\ref{Eq:CoeffDet1}) can be rewritten as
$\langle \lmp ||| B^{(02)} ||| \lm \rangle \sum_{\kappa} c_{\kappa}(L) 
\langle \lm \kappa L ; (02) 0 || \lmp \kappa' L \rangle =0$, where 
$\langle \_ ; \_ || \_ \rangle$ denotes a reduced Wigner coupling coefficient
for SU(3).
Since the triple-reduced matrix element of $B^{(02)}$ is generally nonzero, we 
obtain the following conditions:
\bea
\sum_{\kappa=1}^{\kappa^{max}_L \lm} c_{\kappa}(L) 
\langle \lmp \kappa' L ; (20) 0 || \lm \kappa L \rangle &=& 0  \;\;\;\;\;\; 
(\kappa' = 1,\ldots, \kappa^{max}_L \lm - 1) \;\; .
\eea
Note that the matrix elements of the symplectic generators are not relevant
for the determination of the $c_{\kappa}(L)$, and the SU(3) Wigner coefficients,
$\langle \_ ; \_ || \_ \rangle$, can be evaluated numerically~\cite{Akiyama73}
or analytically~\cite{Hecht90}. 

To illustrate the procedure outlined above, we consider the case of $^{12}$C.
The leading irrep for the nucleus is $\lms$=(0,4), thus the pure states 
belong to $\lm$=(0,4) at 0$\hbo$, $\lm$=(2,4) at 2$\hbo$, $\lm$=(4,4) at 4$\hbo$, 
etc.
At 0$\hbo$, all states ($L$=0,2,4) are unmixed.
At 2$\hbo$, the possible $L$-values are 0, $2^2$, 3, $4^2$, 5, 6, and we have 
$\Delta_{L=0}(2\hbo)$=0 and $\Delta_L(2\hbo)$=1 for $L$=2,3,4,5,6.
Since the values $L$=3,5,6 occur only once ($\kappa_L^{max}[(2,4)]$=1), the
associated states are pure ($c_1(L)$=1.0).
For $L$=2,4, for which $\kappa_L^{max}[(2,4)]$=2, the appropriate coefficients 
$c_{\kappa}(L)$ may be determined from the requirements:
\bea
c_1(L) \langle (0,4)L ; (2,0)0 || (2,4)1 L \rangle 
&+& c_2(L) \langle (0,4)L ; (2,0)0 || (2,4)2 L \rangle = 0  \;\; ,  \nn \\
|c_1(L)|^2 &+& |c_2(L)|^2 = 1   \;\; . 
\eea 
For $L$=2, one finds $ \langle (0,4)2 ; (2,0)0 || (2,4) \kappa 2 \rangle =$ 
-0.85280 (0.05372) for $\kappa=1$ ($\kappa=2$)~\cite{Akiyama73}, and thus
$|\phi_{2M}(2\hbo) \rangle =$ 0.063 $|2\hbo (2,4) 1 2 M \rangle +$
0.998 $|2\hbo (2,4) 2 2 M \rangle$.
Similarly, for $L$=4, one obtains $\langle (0,4)4 ; (2,0)0 || (2,4) \kappa 4 \rangle =$
-0.75107 (0.23440) for  $\kappa=1$ ($\kappa=2$)~\cite{Akiyama73}, and therefore 
$c_1(4) =$ 0.298 and $c_2(4) =$ 0.955.
Analogously, one can proceed for the 4$\hbo$ level.
There are, for instance, three $L$=4 states, one of which is pure.
One finds: $|\phi_{4M}(4\hbo) \rangle =$ -0.637 $|4\hbo (4,4) 1 4 M \rangle +$ 
0.761 $|4\hbo (4,4) 2 4 M \rangle -$ 0.124 $|2\hbo (4,4) 3 4 M \rangle$, and 
similarly for the other states.

For a nucleus with $\lms=(\la,0)$, $\la > 2$, pure states with $\lmp=(\la-2,2)$,
$L=2,3,\ldots,\la-1$, exist at 2$\hbo$ according to the proof given in 
Section~\ref{Sec:PDSHamiltonian}.
The odd-angular momentum values, $L=3,5,\ldots,\la-1$, occur only once 
($\kappa = 1$) and the associated states are pure.
The even-$L$ values, on the other hand, occur twice, with $\kappa = 1$ or 2,
corresponding to Vergados labels 0 and 2, respectively.
Since $\langle (\la,0) L ; (2,0) 0 || (\la-2,2) \kappa L \rangle$ 
= $[2(\la + 1)^2 - L(L+1)]^{1/2}/[3\la (\la + 1)]^{1/2}$ for $\kappa = 1$ and
0 for $\kappa = 2$~\cite{Vergados}, it follows that $c_{\kappa}(L) = 0$ (1.0) 
for $\kappa = 1$ ($\kappa = 2$).
Consequently, the pure K=2 band at 2$\hbo$ consists of states with $\lmp=(\la-2,2)$,
$\kappa = 1$ (2) for odd (even) $L$ values, i.e. $\kappa = \kappa_L^{max}(\la-2,2)$.
An example for such a nucleus is given in Section~\ref{Subs:ApplNe20}, where the
$^{20}$Ne system is discussed.

Having constructed the solvable eigenstates of the PDS Hamiltonian $H(\beta_0,\beta_2)$, 
Eq.~(\ref{Eq:Hpds}), we can now give analytic expressions for their energies.
We have $E(N=0) = 0$ for the 0$\hbo$ level, and  
\bea
E(N) &=& \beta_2 \frac{N}{3} ( N_{\s} - \la_{\s} + \mu_{\s} - 6 + \frac{3}{2} N )
     \;\;\;\;  (\la_{\s} > \mu_{\s}) \nonumber \\
E(N) &=& \beta_2 \frac{N}{3} ( N_{\s} + 2 \la_{\s} + \mu_{\s} - 3 + \frac{3}{2} N )
     \;\;\;\;  (\la_{\s} \leq \mu_{\s})
\label{Eq:PDSEnergies}
\eea
for $N > 0$.
For instance, for $N_\s \lms$ = 24.5 (0,4), which corresponds to $^{12}$C,
this yields: $E(N=0) = 0$, $E(2\hbo) = 19 \beta_2$, $E(4\hbo) = 42 \beta_2$, etc.

The partial SU(3) symmetry of $H(\beta_0,\beta_2)$ is converted into partial
dynamical SU(3) symmetry by adding to the Hamiltonian SO(3) rotation terms which 
lead to $L(L+1)$-type splitting but do not affect the wave functions.
The solvable states then form rotational bands and since their wave functions are 
known, one can evaluate the quadrupole transition rates between them: 
\bea
B(E2, L_i \!\! \rightarrow \!\! L_f) &=& e^2 b^4 \left( \frac{Z}{A} \right)^2
\frac{5}{16\pi} \frac{|\langle L_f || Q_2 || L_i \rangle |^2}{2L_i+1} \;\; .
\eea
Here $b=\sqrt{\hbar/m\om}$ is the harmonic oscillator length parameter, $Z$ and
$A$ are the nuclear charge and mass, respectively, and the convention for the reduced
matrix elements is summarized in Appendix A.
For unmixed initial and final states,
$|\phi_{L_i}(N_i) \rangle =$ 
$\sum_{\kappa_i} c_{\kappa_i}(L_i) |N_i \hbo \lmi \kappa_i L_i \rangle$
and $|\phi_{L_f}(N_f) \rangle =$
$\sum_{\kappa_f} c_{\kappa_f}(L_f) |N_f \hbo \lmf \kappa_f L_f \rangle$,
the matrix element of $Q_2$ is given by:
\bea
\lefteqn{\langle \phi_{L_f}(N_f) || Q_2 || \phi_{L_i}(N_i) \rangle =}   
\nn \\
&& \delta_{N_i,N_f} \delta_{\lmi \lmf} (-1)^{\phi_{\mu_i}} 
\sqrt{6 \langle C_{SU3} \rangle [\lmi]}
\nn \\ 
&& \;\;\; \sum_{\kappa_i \kappa_f} c_{\kappa_i}(L_i) c_{\kappa_f}(L_f)
\langle \lmi \kappa_i L_i ; (11) 2 || \lmi \kappa_f L_f \rangle_{\rho=1} 
\nn \\
&& + \delta_{N_i,(N_f+2)}  \sqrt{3} \langle \lmf ||| A^{(20)} ||| \lmi \rangle
\label{Eq:PDSquad} \\
&& \;\;\; \sum_{\kappa_i \kappa_f} c_{\kappa_i}(L_i) c_{\kappa_f}(L_f)
\langle \lmi \kappa_i L_i ; (20) 2 || \lmf \kappa_f L_f \rangle
\nn \\
&& + \delta_{N_i,(N_f-2)}  \sqrt{3} \langle \lmf ||| B^{(02)} ||| \lmi \rangle
\nn \\
&& \;\;\; \sum_{\kappa_i \kappa_f} c_{\kappa_i}(L_i) c_{\kappa_f}(L_f)
\langle \lmi \kappa_i L_i ; (02) 2 || \lmf \kappa_f L_f \rangle  \; ,
\nn 
\eea
where $\phi_{\mu}$ = 0 for $\mu = 0$ and 1 otherwise.

For intraband transitions, the above expression reduces to the first term on the 
right-hand side. 
For interband transitions there are three possibilities:
For transitions from $N\hbo$ to $(N+2)\hbo$, the second term has to be evaluated;
for $N\hbo \rightarrow (N-2)\hbo$ transitions, the third term is required.
For $\la_{\s} \neq 0$, $\m_{\s} \neq 0$, i.e. for triaxially deformed nuclei,
a $N=0 \rightarrow N=0$ transition is possible as well;
in that case the relevant contribution originates from the first term.
For example, for a transition from $L_i = 2$ to $L_f = 0$ in the ground band of
$^{12}$C, $b = 1.668 fm$, $6 \langle \hat{C}_{SU3} \rangle [(0,4)] = 112$, and thus 
$B(E2, 0\hbo \, L_i\!\!=\!\!2 \!\! \rightarrow \!\! 0\hbo \, L_f\!\!=\!\!0) =$ 0.1925 
$e^2 fm^4$ $\times 112/5$ $\times |\langle (0,4) 2 ; (1,1) 2 || (0,4) 0 \rangle|^2$
= 4.31 $e^2 fm^4$ = 2.64 W.u. (which corresponds to 4.65 W.u., when an effective
charge $e^* = 1.327$ is used).

\section{Applications to light nuclei}
\label{Sec:Applications}

To illustrate that the PDS Hamiltonians of Eq.~(\ref{Eq:Hpds}) are physically
relevant, we present applications to prolate, oblate, and triaxially deformed
nuclei.
We compare energy spectra, reduced quadrupole transition rates, and eigenstates of 
\bea
H_{PDS} &=& h(N) + \xi H(\beta_{0}=12,\beta_{2}=18) + \gamma_2 \hat{L}^2
            + \gamma_4 \hat{L}^4
\eea
to those of the symplectic Hamiltonian
\bea
H_{Sp6} &=& \hat{H}_0 - \chi Q_2 \cdot Q_2 + d_2 \hat{L}^2 + d_4 \hat{L}^4 \; .
\eea  
Here the function $h(N)$, which contains the harmonic oscillator term $\hat{H}_0$,
is simply a constant for a given $N\hbo$ excitation. 
We select light, p-shell and ds-shell, nuclei for which a full, three-dimensional
symplectic calculation can be carried out, that is, a limitation to a submodel
of the Sp(6,R) model is not required.
Since we employ Hamiltonians composed solely of Sp(6,R) generators, we restrict the 
model space to one Sp(6,R) irrep (represented by one `cone' in Fig.~\ref{Sp6Cones}).
We include excitations up to 8$\hbo$.

\subsection{The $^{12}$C case}  

\begin{figure}
\epsfysize=2.0 true in
\epsfbox{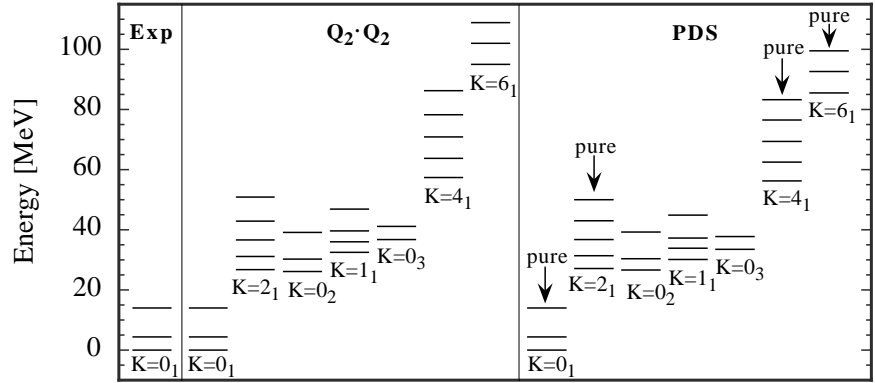}
\vskip 10 pt
\caption{
Energy spectra for $^{12}$C.
Comparison between experimental values (left)~\protect\cite{Ajz90}, results from a
symplectic $8\hbo$ calculation (center) and a PDS calculation (right).
K=$0_1$ indicates the ground band in all three parts of the figure.
In addition, resonance bands dominated by 2$\hbo$ excitations
(K=$2_1,0_2,1_1,0_3$), 4$\hbo$ excitations (K=$4_1$), and 6$\hbo$
excitations (K=$6_1$) are shown for the Sp(6,R) and PDS calculations.
Additional mixed resonance bands (not shown), dominated by 4$\hbo$ and 6$\hbo$ 
excitations, exist for this nucleus.
The angular momenta of the positive parity states in the rotational bands are
$L$=0,2,4,$\ldots$ for K=0 and $L$=K,K+1,K+2, $\ldots$ otherwise.
Bands which consist of pure-SU(3) eigenstates of the PDS Hamiltonian are
indicated.
}
\label{Energies_C12}
\end{figure}

The first nucleus to be considered is $^{12}$C, with four protons and four
neutrons in the valence p-shell. 
This nucleus has previously been studied in the Sp(2,R) submodel of the 
SSM~\cite{Arickx82,Avancini93}.
Here we employ the full, three-dimensional, symplectic shell model. 
The leading Sp(6,R) irrep for this case is $N_{\s} \lms$ $=24.5(0,4)$.
At the 2$\hbo$ level SU(3) irreps $\lm=(2,4)$, (1,3), (0,2) occur, at the 4$\hbo$
level we have $\lm=(0,6)$, (1,4), (2,2)$^2$, (4,4), (3,3), (1,1), (0,0), and so on 
for higher excitations.
The parameters of $H_{Sp6}$ were fitted (simultaneously) to the ground band energies 
and the $2_1^+ \rightarrow 0_1^+$ reduced quadrupole transition strength, for symplectic
model spaces including excitations up to 2$\hbo$, 4$\hbo$, 6$\hbo$, and 8$\hbo$.
The resulting B(E2) strengths are listed in Table~\ref{BE2_C12} and several
low-lying rotational bands are shown in Fig.~\ref{Energies_C12}.
The left part of the figure shows the experimental energies of the ground 
band~\cite{Ajz90}, while the center portion (labeled $Q_2\cdot Q_2$) shows the 
calculated ground band (K=$0_1$), as well as several resonance bands which are 
dominated by 2$\hbo$ excitations (K=$2_1,0_2,1_1,0_3$), 4$\hbo$ excitations 
(K=$4_1$), and 6$\hbo$ excitations (K=$6_1$).
The parameters of $H_{PDS}$ were determined as follows:
$\gamma_2$ and $\gamma_4$ were fixed by the level splittings of the ground band,
$\xi$ was chosen to fit the energy difference between the K=$2_1$ and K=$0_2$
bandheads of the symplectic calculation, and $h(N)$ was adjusted to reproduce 
approximately the relative positions of the K=$0_1,2_1,4_1$, and $6_1$ bandheads.
The resulting spectrum is that shown on the right side of Fig.~\ref{Energies_C12},
labeled PDS.

Since $H_{PDS}$ does not mix states with different $N\hbo$ excitations, the 
B(E2) values obtained in the PDS calculations require an effective charge
$e^*$=1.33 to match the experimental values~\cite{Ajz90} (compare 
Table~\ref{BE2_C12}).
Overall, we find little deviation between the energies and electromagnetic
transition strengths of the two approaches.
A better measure for the level of agreement between the PDS and symplectic
results is given by a comparison of the eigenstates.
According to the proof given in Section~\ref{Sec:PDSHamiltonian}, the Hamiltonian 
$H_{PDS}$ should have sets of solvable, pure-SU(3) eigenstates, which can be 
organized into rotational bands: All 0$\hbo$ states should be pure $\lms = (0,4)$ 
states, and at 2$\hbo$ a rotational band with good SU(3) symmetry $\lm=(2,4)$ and 
$L=2,3,4,5,6$ should exist. 
Similarly, we expect pure-SU(3) bands at 4$\hbo$ with $\lm=(4,4)$ and
$L=4,5,6,7,8$, at 6$\hbo$ with $\lm=(6,4)$ and $L=6,7,8,9,10$, etc. 
An analysis of the PDS eigenstates shows that this is indeed the case.
The associated rotational bands are indicated in Fig.~\ref{Energies_C12}.

\begin{figure}
\hskip 9cm 
\hbox{
\epsfysize=7.0 true in
\epsfbox{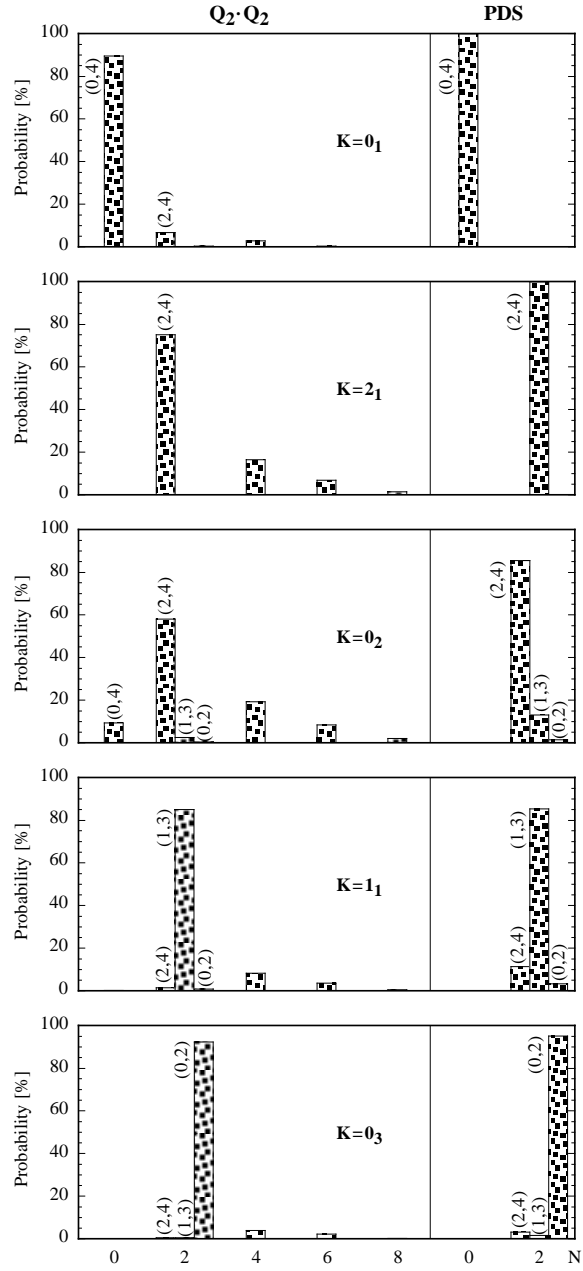}
}
\vskip -5 pt   
\caption{
Decompositions for calculated $L^{\pi}=2^+$ states of $^{12}$C.
Individual contributions from the relevant SU(3) irreps at the 0$\hbo$
and 2$\hbo$ levels are shown for both a symplectic $8\hbo$ calculation
(denoted $Q_{2}\cdot Q_{2}$) and a PDS calculation.
In addition, the total strengths contributed by the $N\hbo$ excitations
for $N>2$ are given for the symplectic case.
}
\label{Decomp_C12_L2}
\end{figure}

\begin{figure}
\hskip 5cm
\hbox{
\epsfysize=6.0 true in
\epsfbox{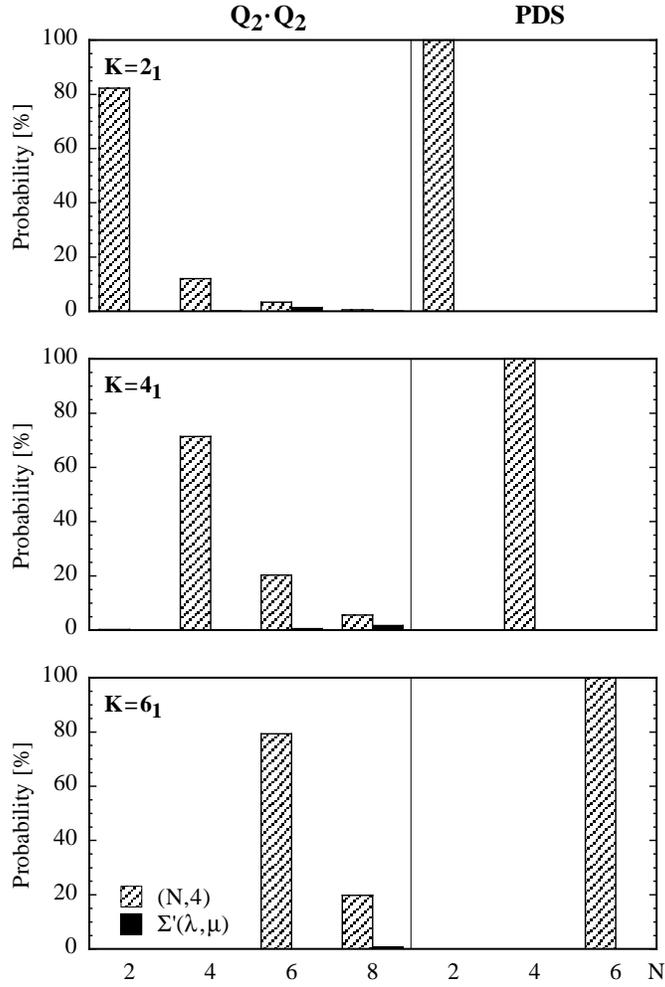}
}
\vskip -35 pt    
\caption{
Decompositions for calculated $L^{\pi}=6^+$ states of $^{12}$C.
The structures shown are representative for the members of the K=$2_1,4_1$,
and $6_1$ rotational bands, respectively.
States of these bands are dominated by $N\hbo$ excited configurations
with $\lm=(N,4)$, $N=2,4,6,8$, in the symplectic scheme and are pure in
the PDS approach.
}
\label{Decomp_C12_L6}
\end{figure}

Figure~\ref{Decomp_C12_L2} shows the decomposition of representative 
($L^{\pi}=2^+$) states of the five lowest rotational bands for the $H_{Sp6}$ 
and $H_{PDS}$ Hamiltonians.
The left side of the figure illustrates the amount of mixing in the wave
functions of the 8$\hbo$ ($Q_{2}\cdot Q_{2}$) calculation: Members of the ground 
band (K=$0_1$) are nearly pure ($\approx 90\%$) 0$\hbo$ states and the resonance 
bands have strong 2$\hbo$ contributions ($\geq 60\%$).
The K=$2_1$, $1_1$, and $0_3$ bands contain admixtures from $N\hbo$ excited
states with $N>2$, while the K=$0_2$ contains admixtures from both the 0$\hbo$
space and from higher oscillator shells.
The relative strengths of the SU(3) irreps within the 2$\hbo$ space are given
as well. 
We find that each rotational band tends to be dominated by one representation,
namely (2,4) for the K=$2_1$ and K=$0_2$ bands, (1,3) for K=$1_1$, and (0,2)
for K=$0_3$, with the other irreps contributing less than 3\%.
The right side of Fig.~\ref{Decomp_C12_L2} shows the structure of the PDS 
eigenstates.
Since the Hamiltonian $H_{PDS}$ does not mix different major oscillator shells,
each eigenstate belongs entirely to one $N\hbo$ level of excitation.
Here the ground band belongs to the 0$\hbo$ space, while the four resonance
bands are pure 2$\hbo$ configurations.
Comparing this with the symplectic results, we observe that the $N\hbo$ level
to which a particular PDS band belongs also dominates the corresponding 
symplectic band.
Furthermore, within this dominant excitation, eigenstates of $H_{Sp6}$ and
$H_{PDS}$ have very similar SU(3) structure, that is, the relative strengths
of the various SU(3) irreps in the symplectic states are approximately
reproduced in the PDS case.
This holds for the K=$0_1$ and K=$2_1$ bands, which are pure in the PDS 
scheme, as well as for the mixed K=$0_2$, $1_1$, and $0_3$ bands.
The above statements are also true for higher $N\hbo$ excitations, as is
illustrated in Fig.~\ref{Decomp_C12_L6} for the $L=6$ states of the 
$N$=2 K=$2_1$, $N$=4 K=$4_1$, and $N$=6 K=$6_1$ bands.
Note also that, in the symplectic case, admixtures from higher shells in the 
$L$=6 wave functions originate predominantly from the `special' irreps
$\lm$ = ($N$,4).

The $^{12}$C example given above nicely illustrates the concept of a partial
dynamical symmetry for a fermionic many-body system.
The pure PDS eigenstates form rotational bands which follow the pattern for
solvable states of family (b),
their energies and E2 transition strengths between them can be evaluated
analytically according to Eqs.~(\ref{Eq:PDSEnergies})--(\ref{Eq:PDSquad}).

\subsection{The $^{20}$Ne case} 
\label{Subs:ApplNe20}    
We now turn to a system with pure PDS eigenstates that follow pattern (a):
The $^{20}$Ne nucleus, with two valence protons and neutrons each, has
previously been described within the symplectic model 
framework~\cite{SymplM,Casta89a,Draayer84,Suzuki87}.
The leading Sp(6,R) irrep for this prolate nucleus is $N_{\s}\lms$ $=48.5(8,0)$.
We expect to find solvable, pure-SU(3) eigenstates of $H_{PDS}$ at 0$\hbo$,
2$\hbo$, and 4$\hbo$.
More specifically, there should be a  K=$0_1$ $L=0,2,4,6,8$ rotational band
with $\lm=$ (8,0) at 0$\hbo$, a  K=$2_1$ $L=2,3,4,5,6,7$ band with $\lm=$
(6,2) at 2$\hbo$, and a K=$4_1$ $L=4,5$ `band' with $\lm=$ (4,4) at 4$\hbo$.
Pure PDS states at higher levels of excitation do not exist.

\begin{figure}
\hskip -1cm
\epsfysize=2.5 true in
\epsfbox{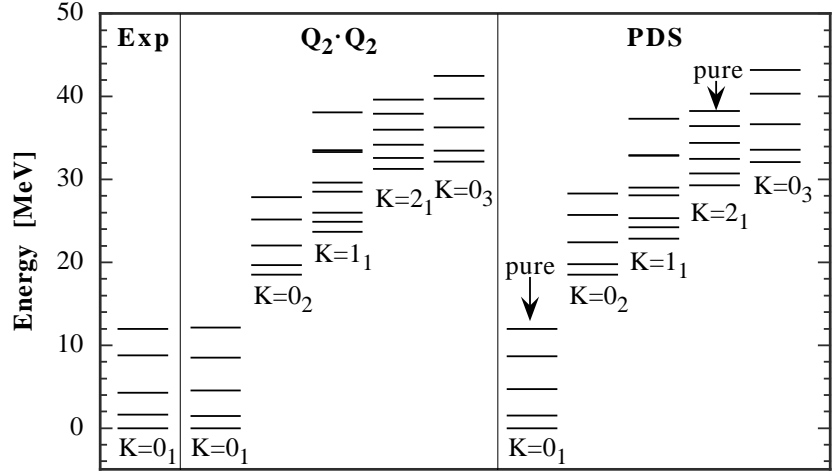}
\vskip 10 pt  
\caption{
Energy spectra for $^{20}$Ne.
Experimental ground band (K=$0_1$) energies~\protect\cite{Tilley98} are shown on
the left, while theoretical results for both the ground band and 2$\hbo$
resonances (K=$0_2,1_1,2_1,0_3$) are given in the center and on the right,
for a symplectic $8\hbo$ and a PDS calculation, respectively.
Rotational bands which consist of pure eigenstates of the PDS Hamiltonian
are indicated.
}
\label{Energies_Ne20}
\end{figure}

As in the $^{12}$C case, we compare the eigenstates of $H_{PDS}$ to those of 
the symplectic Hamiltonian $H_{Sp6}$.
Least squares fits to measured energies and B(E2) values of the ground
band of $^{20}$Ne~\cite{Tilley98} were carried out for 2$\hbo$, 4$\hbo$, 
6$\hbo$, and 8$\hbo$ symplectic model spaces.
The resulting energies and transition rates converge to values which
agree with the data, as is illustrated in Fig.~\ref{Energies_Ne20} and
Table~\ref{BE2_Ne20_GB}.
The parameters $\gamma_2$ and $\gamma_4$ in $H_{PDS}$ were determined by
the energy splitting between states of the ground band,
$\xi$ was adjusted to reproduce the relative positions of the 2$\hbo$
resonance bandheads and $h(N)$ was fixed by the energy difference
$[E(0_2^+) - E(0_1^+)]$.
Figure~\ref{Energies_Ne20} and Table~\ref{BE2_Ne20_GB} demonstrate the level
of agreement between the PDS and symplectic results.
         
\begin{figure}
\hskip 11cm
\hbox{
\epsfysize=7.0 true in
\epsfbox{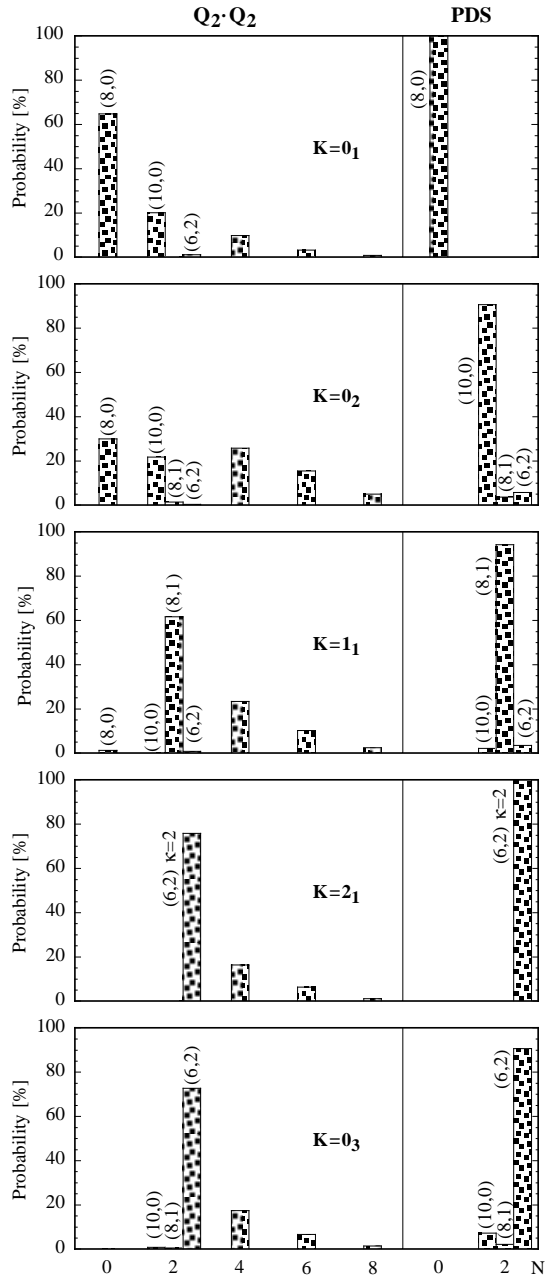}
}
\vskip -5 pt  
\caption{
Decompositions for calculated $L^{\pi}=2^+$ states of $^{20}$Ne.
Individual contributions from the SU(3) irreps at the 0$\hbo$
and 2$\hbo$ levels are shown for both a symplectic $8\hbo$ calculation
(left side) and a PDS calculation (right side).
For the symplectic approach the summed contributions from SU(3) irreps 
at higher ($N>2$) excitations are given as well.
}
\label{Decomp_Ne20}
\end{figure}

An analysis of the structure of the ground and resonance bands reveals
the amount of mixing in the 8$\hbo$ symplectic ($Q_{2}\cdot Q_{2}$)
wave functions.
Figure~\ref{Decomp_Ne20} shows the decomposition for representative 
($L^{\pi}=2^{+}$) states of the five lowest rotational bands.
Ground band (K=$0_1$) states are found to have a strong $0\hbo$ component
($\geq 64\%$), and three of the four resonance bands are clearly
dominated ($\geq 60\%$) by $2\hbo$ configurations.
States of the first resonance band (K=$0_2$), however, contain significant
contributions from all but the highest $N\hbo$ excitations.
The relative strengths of the SU(3) irreps within the $2\hbo$ space are
shown as well:  as in the $^{12}$C case, states are found to be dominated 
by one representation [(10,0) for the K=$0_2$ band, (8,1) for K=$1_1$, 
(6,2)$\kappa=2$ for K=$2_1$, and (6,2)$\kappa=1$ for K=$0_3$ here], while 
the other irreps contribute only a few percent. 
Such trends are present also in the more realistic symplectic calculations 
of Suzuki~\cite{Suzuki87}.                               
                                                        
As expected, $H_{PDS}$ has families of pure-SU(3) eigenstates which can be
organized into rotational bands, Fig.~\ref{Energies_Ne20}.
The ground band belongs entirely to $N=0$, $\lm=(8,0)$, and all states of
the K=$2_1$ band have quantum labels $N=2$, $\lm=(6,2)$, $\kappa=2$.
The K=$4_1$ band at 4$\hbo$ is not shown.
A comparison with the symplectic case shows that the $N\hbo$ level to
which a particular PDS band belongs is also dominant in the corresponding
symplectic band, Fig.~\ref{Decomp_Ne20}.
As before, within this dominant excitation, eigenstates of $H_{PDS}$
and $H_{Sp6}$ have similar SU(3) distributions; in particular, both
Hamiltonians favor the same $\lm\kappa$ values.  
Significant differences in the structure of the wave functions appear,
however, for the K=$0_2$ resonance band.
In the $8 \hbo$ symplectic calculation, this band contains almost equal
contributions from the $0 \hbo$, $2 \hbo$, and $4 \hbo$ levels, with
additional admixtures of $6 \hbo$ and $8 \hbo$ configurations, while in
the PDS calculation, it belongs entirely to the $2 \hbo$ level.
These structural differences are also evident in the interband transition 
rates, as is illustrated in Table~\ref{BE2_Ne20_Comp}.
Whereas the intraband B(E2) strengths are approximately equal in both 
approaches, we observe that the interband rates differ by a factor of
2-3 in most cases.
These differences reflect the action of the inter-shell coupling terms
that are present in the quadrupole-quadrupole interaction of 
Eq.~(\ref{Eq:QQ}), but do not occur in the PDS Hamiltonian.
Increasing the strength $\chi$ of $Q_2 \cdot Q_2$ in $H_{Sp6}$
will also spread the other resonance bands over many $N\hbo$ excitations.
The K=$2_1$ band (which is pure in the PDS scheme) is found to resist
this spreading more strongly than the other resonances.
For physically relevant values of $\chi$, the low-lying bands have
the structure shown in Fig.~\ref{Decomp_Ne20}.
                                                
\subsection{The $^{24}$Mg case}      
The final example to be considered here involves the triaxially deformed
nucleus $^{24}$Mg, which has been the subject of several symplectic
model studies~\cite{Rosen84,Casta89a,Reske84,Escher99PRL}.
With four valence protons and neutrons in the ds-shell each, and a
leading Sp(6,R) irrep $N_{\s}\lms$ $=62.5(8,4)$, this system is the most
complicated one to be investigated here.
Since both $\la_{\s} \neq 0$ and $\mu_{\s} \neq 0$, the symplectic
Hilbert space has a very rich structure.
The (8,4) representation at 0$\hbo$ contains three rotational bands:
a K=0 band with $L=0,2,4,6,8$, a K=2 band with $L=2,3,\ldots,10$,
and a K=4 band with $L=4,5,\ldots,12$.
At the 2$\hbo$ level, there are six possible SU(3) irreps, (10,4), (8,5),
(6,6), (9,3), (7,4), and (8,2), which contain a total of four K=0, two K=1,
four K=2, two K=3, three K=4, one K=5, one K=6, $\ldots$ bands.
At the 4$\hbo$ level, there are 15 different SU(3) irreps, at 6$\hbo$,
there are 25, etc.
Accordingly, the number of states for a given angular momentum value $L$
increases dramatically with the inclusion of higher excitations.
This is illustrated in Table~\ref{Dim_Mg24} for $L=0,1,2,\ldots,8$.

Since the interactions in $H_{Sp6}$ do not distinguish different
$\kappa$-multiplicities, it becomes necessary to make use of the 
integrity basis operators $\hat{X}_3$ and $\hat{X}_4$ discussed in 
Section~\ref{Sec:SSM}, which allow us to reproduce the experimentally 
observed K-band splitting in the spectrum of $^{24}$Mg. 
Using the Hamiltonian 
\bea
H_{Sp6}' = H_{Sp6} + c_3 \hat{X}_3 + c_4 \hat{X}_4 \; ,
\eea 
we carried out least squares fits to measured energies and B(E2) values for 
2$\hbo$, 4$\hbo$, and 6$\hbo$ symplectic spaces.

\begin{figure}
\hbox{
\epsfysize=7.0 true in
\epsfbox{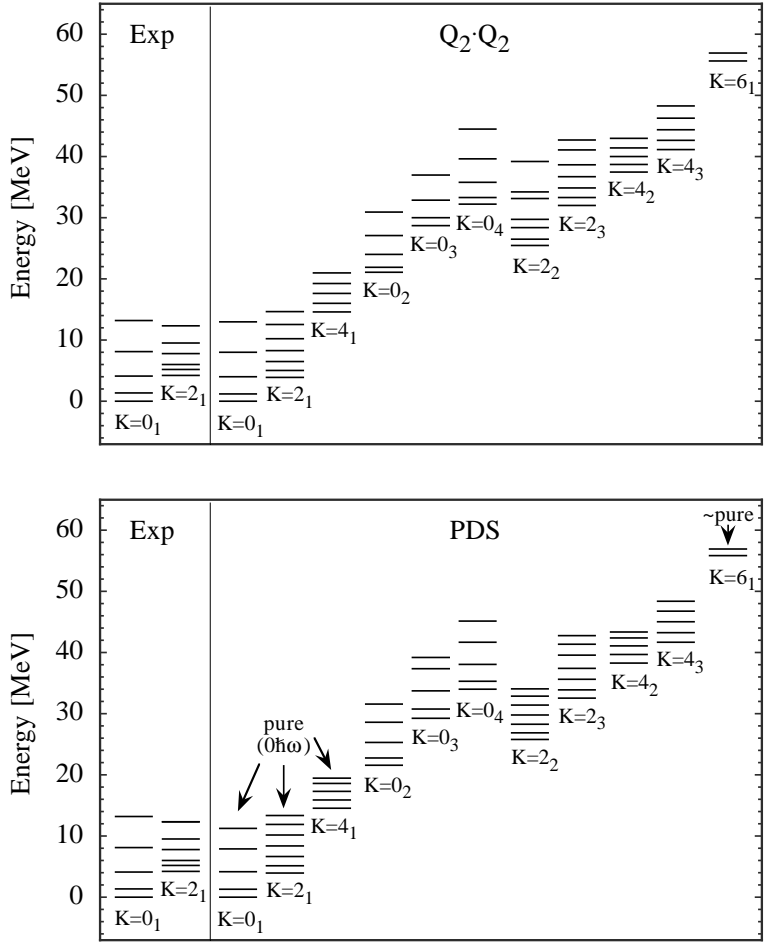}
}
\vskip -80 pt
\caption{
Energy spectra for $^{24}$Mg.
Energies from a PDS calculation (bottom) are compared to symplectic results
(top).
Both 0$\hbo$-dominated bands (K=$0_1,2_1,4_1$) and some 2$\hbo$ resonance
bands (K=$0_2,0_3,0_4,2_2,2_3,4_2,4_3,6_1$) are shown.
The K=$0_1,2_1,4_1$ ($6_1$) states are pure (approximately pure) in the PDS
scheme.
Experimental values for the ground and $\gamma$-band energies, taken
from Refs.~\protect\cite{Branf75,Endt93}, are given on the left.
}
\label{Energies_Mg24}
\end{figure}

Figure~\ref{Energies_Mg24} (top) displays the energies obtained with the 
6$\hbo$ calculation (right part of the figure) in comparison with the 
experimental values~\cite{Branf75,Endt93} (left side).
In addition to the ground (K=$0_1$) and $\gamma$ (K=$2_1$) bands, the
calculated K=$4_1$ band, which is dominated by 0$\hbo$ configurations, and
several low-lying symplectic K=$0,2,4$, and 6 bands, which are dominantly 
2$\hbo$ resonances, are shown.
Table~\ref{BE2_Mg24_GG} lists various B(E2) transition rates between the 
low-lying states of $^{24}$Mg.
We find that the results of the symplectic calculations are in good agreement 
with the data
\footnote{
Note that we have used the experimental B(E2) values from Ref.~\cite{Branf75}, 
which contains a more complete list of B(E2) data than the compilation by
Endt~\cite{Endt93}.  The latter gives values of 20.6$\pm$0.4 W.u.,
35$\pm$5 W.u., and 37$\pm$12 W.u. for the first three transitions listed
in Table~\ref{BE2_Mg24_GG}. 
Fitting the symplectic Hamiltonian parameters to reproduce the values of
Ref.~\cite{Endt93} gives results very similar to the ones presented here
and does not alter our conclusions. 
}.
Specifically, the $\gamma$-band is correctly located and nearly all the
calculated intraband and interband transition rates fall, without the use
of an effective charge, within experimental uncertainties.
The 4$\hbo$ results are better than the 2$\hbo$ results, with the 6$\hbo$
calculation yielding only moderate improvements.

In analogy with the symplectic case, we include terms $\hat{X}_3$ and 
$\hat{X}_4$ in the PDS Hamiltonian: 
\bea
H_{PDS}' = H_{PDS} + c_3 \hat{X}_3 + c_4 \hat{X}_4 \; .
\eea
As we will see below, the introduction of these extra terms breaks the partial 
symmetry.
We fixed $c_3$ and $c_4$ at the values that were used in the 6$\hbo$
symplectic calculation, determined $\gamma_2$ and $\gamma_4$ from the level
splittings in the K=$0_1$ ground band, and adjusted $\xi$ so as to reproduce
the relative positions of selected 2$\hbo$ bandhead states (we focused on
the lowest three K=0 bands and the first K=6 band).
Then $h(N)$ was chosen to reproduce approximately the positions of the 2$\hbo$
resonances relative to the ground and $\gamma$ bands.

We obtain an energy spectrum which agrees well with the results of the symplectic
calculation, as is shown in Fig.~\ref{Energies_Mg24}.
The B(E2) strengths for the ground and $\gamma$-bands, rescaled by an effective
charge $e^*$=1.75, are given in Table~\ref{BE2_Mg24_GG}.
We find good agreement between the PDS and symplectic calculations for the 
intraband transitions, whereas there are larger deviations in the interband
rates.

\begin{figure}
\hskip -1.0cm
\hbox{
\epsfysize=5.4 true in
\epsfbox{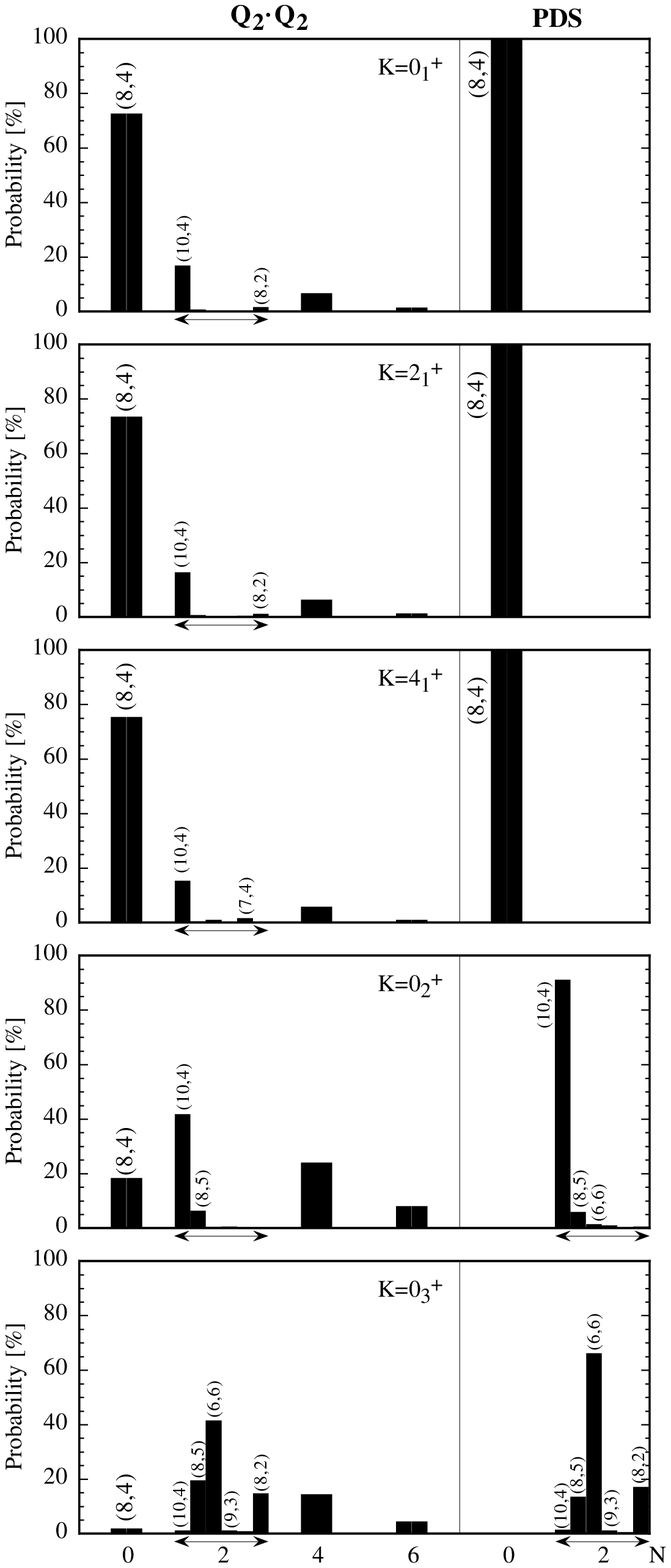}
}
\hskip -4cm 
\hbox{
\epsfysize=5.4 true in
\epsfbox{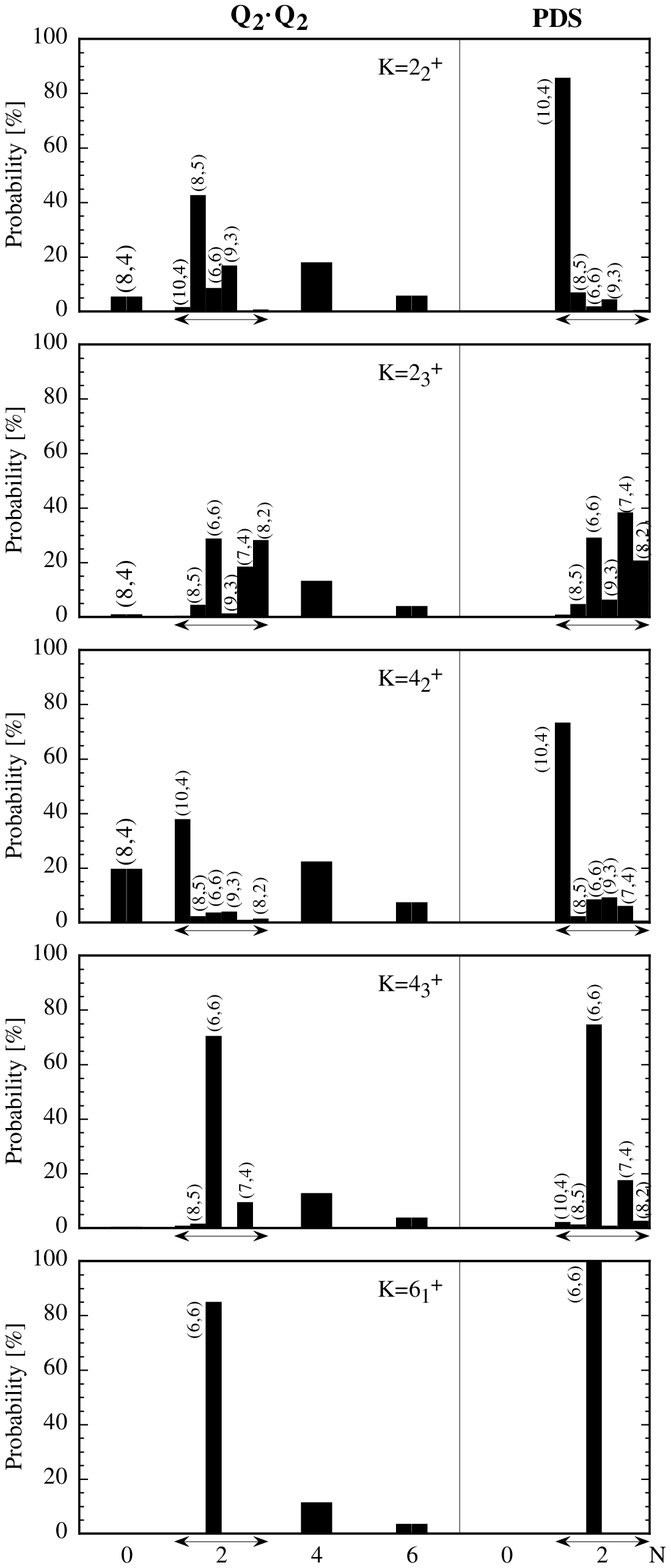}
}
\caption{
Decompositions for calculated $L^{\pi}=6^+$ states of $^{24}$Mg.
Eigenstates resulting from the symplectic 6$\hbo$ calculation are decomposed
into their 0$\hbo$, 2$\hbo$, 4$\hbo$, and 6$\hbo$ components (denoted by
$Q_{2}\cdot Q_{2}$ in the figure).
At the 0$\hbo$ and 2$\hbo$ levels, contributions from the individual SU(3)
irreps are shown, for higher excitations ($N>2$) only the summed strengths
are given.
Eigenstates of the PDS Hamiltonian belong entirely to one $N\hbo$ level of
excitation, here 0$\hbo$ or 2$\hbo$.
Contributions from the individual SU(3) irreps at these levels are shown.
Members of the K=$0_1,2_1,4_1$ bands are pure in the PDS scheme, and K=$6_1$
states are very nearly ($>99\%$) pure.
}
\label{Decomp_Mg24}
\end{figure}

According to the proof given in Section~\ref{Sec:PDSHamiltonian}, the three 
rotational bands at 0$\hbo$ should be pure in the PDS scheme, and at 2$\hbo$ 
there should be a (short) rotational K=6 band with $L$=6,7, which belongs 
entirely to the $\lm=(6,6)$ representation.
We find that the 0$\hbo$ states are indeed pure, but the K=6 $L=6,7$ band
has small admixtures ($<1\%$) from 2$\hbo$ irreps other than $\lm=(6,6)$,
thus indicating that $H_{PDS}'$, unlike $H_{PDS}$, is not an exact partial 
dynamical symmetry Hamiltonian, due to the presence of the K-band splitting 
terms $\hat{X}_3$ and $\hat{X}_4$.
This can be understood as follows:
While $\hat{X}_3$ and $\hat{X}_4$ cannot mix different SU(3) irreps, their 
eigenstates involve particular linear combinations of different $\kappa$ values.
Since the operators $\hat{X}_3$ and $\hat{X}_4$ do not commute with $B_0$, 
these linear combinations differ from configurations resulting from the PDS 
requirement $B_0 |\phi \rangle = 0$.
Fortunately, a very small amount of symmetry-breaking suffices to fit the 
relative positions of the ground and $\gamma$-bands, as can be inferred from
the eigenstate decompositions plotted in Fig.~\ref{Decomp_Mg24}.
Shown are the decompostions of the $L=6$ states associated with the calculated 
$H_{Sp6}'$ and $H_{PDS}'$ spectra.
More specifically, we have plotted the contributions from the SU(3) irreps at
0$\hbo$ and 2$\hbo$, as well as the (summed) contributions from 4$\hbo$ and
6$\hbo$ excitations.

As in the previous examples, we observe that the eigenstates of both 
Hamiltonians have very similar structures:  For a given state, the same
$N\hbo$ level of excitation is dominant in both calculations and, moreover,
within this dominant excitation, we find similar SU(3) distributions.
The structural differences that do exist are, again, reflected in the very
sensitive interband transition rates, as can be seen in 
Table~\ref{BE2_Mg24_Comp}.

\section{Comparison of partial symmetries in bosonic and fermionic many-body 
systems}
\label{Sec:IBMComp}

Partial dynamical symmetries were first studied in the Interacting Boson 
Model (IBM) of nuclei~\cite{IBMBooks}. In~\cite{Leviatan96a}, the following IBM
Hamiltonian was used to reproduce measured energies and E2 rates of $^{168}$Er:
\bea
H_{IBM}(h_0,h_2) &=& h_0 P^{\dagger}_0 P_0 + h_2 P^{\dagger}_2 \cdot \tilde{P}_2 \; ,
\label{Eq:IBMHam}
\eea
where $h_0,h_2$ are arbitrary parameters and $P^{\dagger}_L$, $L=0$ and 2, 
are boson pair operators:
\bea
P^{\dagger}_0 &=& d^{\dagger} \cdot d^{\dagger} - 2 (s^{\dagger})^2  \; , \nn \\
P^{\dagger}_{2\mu} &=& 2s^{\dagger}d^{\dagger}_{\mu} 
                    + \sqrt{7} (d^{\dagger} d^{\dagger})^{(2)}_{\mu}  \; .
\eea
The creation operators $s^{\dagger}$ and $d^{\dagger}_{\mu}$ ($\mu=0,\pm1,\pm2$)
denote a monopole boson with angular momentum and parity $J^{\pi}=0^+$, and a 
quadrupole boson with $J^{\pi}=2^+$, respectively.  
They represent correlated valence nucleon pairs and are the basic building 
blocks of the IBM.
The pair operators $P^{\dagger}_0$ and $P^{\dagger}_{2\mu}$ are components of
a $\lm=(0,2)$ SU(3) tensor, and their Hermitean adjoints, $P_0$ and 
$\tilde{P}_{2\mu}=(-1)^{\mu} P_{2,-\mu}$, are characterized by $\lm=(2,0)$.

It can be shown that for $h_2=h_0$, the Hamiltonian of Eq.~(\ref{Eq:IBMHam}) 
becomes a SU(3) scalar (related to the Casimir operator of $SU(3)$) and for 
$h_2=-h_0/5$, it transforms as a $\lm=(2,2)$ SU(3) tensor component.
In general, $H_{IBM}(h_0,h_2)$ is therefore not a SU(3) scalar, nevertheless
it turns out that it always has an exact zero-energy eigenstate, denoted in what
follows by $|c;N\rangle$, where the integer $N$ gives the total number of bosons 
in the system.
The state $|c;N\rangle$ describes a condensate of bosons and can be written as 
\bea
|c;N\rangle &=& \frac{1}{\sqrt{N!}} 
  \left[ ( s^{\dagger} + \sqrt{2} d^{\dagger}_0 )/\sqrt{3} \right]^N | 0 \rangle \; .
\label{Eq:BosCond}
\eea 
It is the lowest weight state in the SU(3) irrep $\lm=(2N,0)$ and serves as
an intrinsic state for the SU(3) ground band.
The rotational members of the ground band with good angular momentum $L$ are
obtained by projection from $|c;N\rangle$.
Moreover, one finds that states of the form
\bea
| k \rangle \propto (P^{\dagger}_{22})^k |c;N\rangle
\eea
are eigenstates of $H_{IBM}(h_0,h_2)$ with eigenvalues $E_k=6h_2(2N+1-2k)k$
and good SU(3) symmetry $(2N-4k,2k)$, where $2k \leq N$.
They are lowest weight states in these representations and serve as intrinsic
states representing $\gamma^k$ bands with angular momentum projection $K=2k$
along the symmetry axis.

Since $H_{IBM}(h_0,h_2)$ is rotationally invariant, it follows that states of
good $L$ projected from $ |k = 0 \rangle = |c;N\rangle$ and $| k \rangle$, 
$k > 0$, are also eigenstates with energy $E_k$ and SU(3) symmetry $(2N-4k,2k)$.
The projected states span the entire $(2N,0)$ representation for $k=0$, but
only part of the corresponding irrep for $k>0$.
There are other excited states which do not preserve the SU(3) symmetry and 
therefore contain a mixture of SU(3) irreps, including the `special' irreps
$(2N-4k,2k)$.
Since $H_{IBM}(h_0,h_2)$ is not a SU(3) scalar, but possesses a subset of 
solvable eigenstates with good SU(3) symmetry, it is a partial symmetry
Hamiltonian.
Adding $\hat{L}^2$, the Casimir operator of SO(3), to $H_{IBM}(h_0,h_2)$ converts 
the partial symmetry to a partial dynamical symmetry and contributes a $L(L+1)$
splitting, but does not affect the wave functions.

The boson and fermion Hamiltonians, $H_{IBM}(h_0,h_2)$ of Eq.~(\ref{Eq:IBMHam})
and $H(\beta_0,\beta_2)$ of Eq.~(\ref{Eq:Hpds}), have several features in 
common:
Both display partial SU(3) symmetry, they are constructed to be rotationally 
invariant functions of $\lm=(2,0)$ and $\lm=(0,2)$ SU(3) tensor operators, and 
SU(3) tensor decompositions show that both contain $\lm=(0,0)$ and (2,2) terms
only.
$H_{IBM}(h_0,h_2)$, as well as $H(\beta_0,\beta_2)$, has solvable pure-SU(3)
eigenstates, which can be organized into rotational bands; the degeneracies
within these bands are lifted by adding the SO(3) term $\hat{L}^2$ to the
Hamiltonian.
The ground bands are pure in both cases, and higher-energy pure bands coexist
with mixed-symmetry states.

There are several significant differences between the bosonic and fermionic
PDS Hamiltonians, however.
For example, the ground band of the Hamiltonian $H_{IBM}(h_0,h_2)$, 
Eq.~(\ref{Eq:IBMHam}), is characterized by $\lm = (2N,0)$, i.e., it describes
an axially-symmetric prolate nucleus.
Is is also possible to find an IBM Hamiltonian with partial SU(3) symmetry
for an oblate nucleus.
It can be shown that these two cases exhaust all possibilities for partial
SU(3) symmetry with a two-body Hamiltonian in the IBM-1 with one type of monopole 
and quadrupole bosons.
In contrast, the fermionic Hamiltonians considered here can accomodate ground
bands of prolate [$(\la_{\s},0)$], oblate [$(0,\mu_{\s})$], and triaxial
[$(\la_{\s},\mu_{\s})$ with $\la_{\s} \neq 0$, $\mu_{\s} \neq 0$] shapes.

Another difference between the fermionic and the bosonic PDS cases discussed
here lies in the physical interpretation of the excited solvable bands.
While these bands represent $\gamma$, double-$\gamma$, etc.\/ excitations in 
the IBM, they correspond to giant monopole and quadrupole resonances in the 
fermion case.               

Furthermore, whereas the pure eigenstates of $H_{IBM}(h_0,h_2)$ can be
generated by repeated action of the boson pair operator $P^{\dagger}_{22}$
on the boson condensate and subsequent angular momentum projection, a similar
straightforward construction process for the special eigenstates of 
$H(\beta_0,\beta_2)$ has not been identified yet.
The situation seems to be more complicated in the fermion case, which is
also reflected in the fact that $H(\beta_0,\beta_2)$ has two possible families
of pure eigenstates, one finite, the other infinite.
The association of the special states to one or the other family depends on
the 0$\hbo$ symplectic starting configuration.

The comparison of partial dynamical symmetries in bosonic and fermionic systems
above illustrates that, in spite of similar algebraic structures of the associated
Hamiltonians, two given systems with partial symmetries may exhibit not only
different physical interpretations, but also different systematic features
and different mechanisms for generating the partial symmetries in question.

\section{Summary and Conclusions}
\label{Sec:Summary}

The fundamental concept underlying algebraic theories in quantum physics is 
that of an exact or dynamical symmetry.
Realistic quantum systems, however, often require the associated symmetry
to be broken in order to allow for a proper description of some observed
basic features.
Partial dynamical symmetry describes an intermediate situation in which 
some eigenstates exhibit a symmetry which the associated Hamiltonian 
does not share.
The objective of this approach is to remove undesired constraints from the
theory while preserving the useful aspects of a dynamical symmetry, such
as solvability, for a subset of eigenstates.

We have presented an example of a partial dynamical symmetry in
an interacting many-fermion system.
In the framework of the symplectic shell model, we have constructed a 
family of rotationally invariant Hamiltonians with partial SU(3) symmetry.
We have demonstrated that the PDS Hamiltonians are closely related to the
deformation-inducing quadrupole-quadrupole interaction and break SU(3)
symmetry, but still possess a subset of `special' solvable eigenstates
which respect the symmetry.
The construction process for these special states was outlined and
analytic expressions for their energies and for E2 transition
rates between them were given.

To illustrate that the PDS Hamiltonians introduced here are physically
relevant, we have presented applications to oblate, prolate, and
triaxially deformed nuclei.
Specifically, we have compared the energy spectra, reduced quadrupole
transition strengths, and eigenstate structures of the partial symmetry
Hamiltonians to those of a symplectic shell model Hamiltonian containing 
a realistic quadrupole-quadrupole interaction.
Although the PDS Hamiltonians cannot account for intershell correlations, 
we have observed that various features of the quadrupole-quadrupole interaction 
are reproduced with a particular parameterization of the partial symmetry 
Hamiltonians.
PDS eigenfunctions do not contain admixtures from different $N\hbo$
configurations, but belong entirely to one level of excitation.
We have found that, for reasonable interaction parameters, the $N\hbo$ level
to which a particular PDS band belongs is also dominant in the corresponding
band of exact $Q_2 \cdot Q_2$ eigenstates.
Moreover, within this dominant excitation, eigenstates of both Hamiltonians
have similar SU(3) distributions.
Structural differences, nevertheless, do arise and are reflected in the very 
sensitive interband transition rates.       
Overall, however, we may conclude that PDS eigenstates approximately 
reproduce the structure of the exact $Q_2 \cdot Q_2$ eigenstates, for both
ground and most resonance bands.

The notion of partial dynamical symmetries extends and complements the
familiar concepts of exact and dynamical symmetries.
It is applicable when a subset of physical states exhibit a symmetry which
does not arise from the invariance properties of the relevant Hamiltonian.
Recent studies, including the one presented here, show that partial
symmetries may indeed be realized in various quantum systems.
This indicates that PDS is not a mere mathematical concept, but may serve
as a practical tool in realistic applications of algebraic methods to 
physical systems.

\section*{Acknowledgements}

The authors appreciate valuable discussions with D.\/J.\/ Rowe,
J.\/P.\/ Elliott, and G.\/ Rosensteel during a visit to the Institute
for Nuclear Theory at the University of Washington.
This work is supported in part by the Israel Science Foundation.

\section*{Appendix A: SU(3) Wigner coefficients and Wigner-Eckart theorem}
\label{Sec:AppA}

If $\al$ represents a set of labels used to distinguish orthonormal basis
states within a given irreducible SU(3) representation $\lm$, the Wigner
coefficients
$\langle \, \lme \al_1; \lmz \al_2 \, | \, \lm \al \, \rangle_\rho$ are
defined as the
elements of a unitary transformation between coupled and uncoupled orthonormal
irreps of SU(3) in the $\al$-scheme~\cite{JPD73a}:
\bea
| \lm \al \, \rangle_{\rho}
= \sum_{\al_1 \al_2}  \langle \, \lme \al_1; \lmz \al_2 \, | \, \lm \al \,
\rangle_{\rho} \,
| \,\lme \al_1 \, \rangle | \, \lmz \al_2 \, \rangle,
\eea
and the inverse transformation is given by:
\bea
| \, \lme \al_1 \rangle | \, \lmz \al_2 \, \rangle
&=& \sum_{\rho \lm \al} \langle \, \lme \al_1 ; \lmz \al_2 \, | \, \lm \al \,
\rangle_\rho \: | \lm \al \, \rangle_{\rho} \: .
\eea
Here $\al = \epsilon \Lambda M_{\Lambda}$ for the SU(3) $\supset$ SU(2)
$\otimes$ U(1) (canonical) group chain and $\al = \k l m$ for the SU(3) 
$\supset$ SO(3) reduction employed in this work.
The subgroup chains impose certain restrictions on the above couplings, for
example the usual angular momentum coupling rules,
$l=l_1+l_2$, $\ldots$, $|l_1 - l_2|$, and $m=m_1+m_2$ apply for the chain
containing SO(3).

The outer multiplicity label $\rho = 1,2, \ldots, \rho_{max}$ is used to
distinguish multiple occurrences of a given
$\lm$ in the direct product $\lme \times \lmz$: $\rho = 1,2,\ldots,\rho_{max}$, 
where $\rho_{max}$ denotes the number of possible couplings $\lme \times \lmz$, 
and the possible $\lm$ irreps in the product can be obtained by coupling the
appropriate Young diagrams~\cite{Hamermesh}.
O'Reilly~\cite{OReilly82} determines a closed formula for the decomposition of 
the outer product $\lme \times \lmz$ of SU(3) irreps for 
arbitrary positive integers $\la_i, \mu_i$ and derives necessary and sufficient 
conditions for a SU(3) irrep $\lm$ to appear as summand in $\lme \times \lmz$.

It is possible to factor out the dependence of the above SU(3) $\supset$
SO(3) Wigner coupling coefficient on the $m$ subgroup label by defining so-called
double-barred or ``reduced'' SU(3) coupling coefficients:
\bea
\lefteqn{ \sutcc{ \lme \k_1 l_1 m_1} { \lmz \k_2 l_2 m_2} { \lm \k l m}_\rho}
\nn \\
&& =  \underbrace{
\sutrcc{\lme \k_1 l_1} {\lmz \k_2 l_2 } { \lm \k l}_\rho }_{\mbox{reduced
Wigner coefficient}}
\, \,
\underbrace{
\cg{l_1}{m_1}{l_2}{m_2}{l}{m}  }_{\mbox{geometric part}}  \; .
\eea
The ``geometric'' part $\langle \_ \: \_ \: | \: \_ \rangle$ is simply a
SU(2) Clebsch-Gordan coefficient.  From the unitarity of the full SU(3) 
Wigner and the ordinary SU(2) Clebsch-Gordan coefficients it follows that 
the double-bar coefficients are also unitary.  With the phase convention 
introduced in Ref.~\cite{JPD73a} they become real, and therefore orthogonal.                
Draayer and Akiyama~\cite{JPD73a} give a prescription for the unique
determination, including the phases, of SU(3) Wigner coefficients and 
derive their relevant conjugation and symmetry properties.    
They furthermore provide a computer code which allows for a numerical 
determination of the coefficients~\cite{Akiyama73}.  
Analytic expressions for Wigner coefficients which are of particular 
interest in p-shell and ds-shell nuclear shell-model calculations are 
tabulated in Ref.~\cite{Hecht65} for the canonical subgroup chain and in
Ref.~\cite{Vergados,Hecht90,Hecht82a} for the SU(3) $\supset$ SO(3) chain.
                                                               
The  Wigner-Eckart theorem for the group SU(2) yields SU(2)-reduced
(double-bar) matrix elements of a SO(3) irreducible tensor 
operator:
\bea
\lefteqn{
\langle \, l_3 m_3 \, | \, T^{l_2 m_2} \, | \, l_1 m_1 \,\rangle}
\nn \\
&& = \langle \, l_1 m_1 ; l_2 m_2 | \, l_3 m_3 \rangle \;
\frac{\langle \, l_3 \,|| \, T^{l_2}\, || l_1 \, \rangle}
{\sqrt{2 l_3 +1}}  \;  .
\label{WESO3}
\eea  
Analogously, the generalized Wigner-Eckart theorem allows one to express
matrix elements of SU(3)
irreducible tensor operators as a sum over $\rho$ of the product of a
$\rho$-dependent generalized
reduced matrix element multiplied by the corresponding Wigner coefficient
\cite{JPD73a}:
\bea
\lefteqn{
\langle \, \lmd \al_3 \, | \, T^{\lmz \al_2} \, | \, \lme \, \al_1 \,\rangle}
\nn \\
&& = \sum_\rho  \mbox{ } \langle \, \lme \, \al_1 ; \lmz \,
\al_2 | \, \lmd \al_3 \rangle_\rho \;
\langle \, \lmd \,|||  \, T^{\lmz}\, ||| \lme \, \rangle_\rho  \;  .
\label{WEcan}
\eea
For more details on SU(3) coupling and recoupling coefficients, see the
compilation in Appendix C of Ref.~\cite{Escher98b} and references therein.

\section*{Appendix B: Matrix elements of relevant operators}
\label{Sec:AppB}  

The calculations presented here require expressions for matrix elements of
the Sp(6,R) generators $\hat{A}^{(20)}$, $\hat{B}^{(02)}$, and $\hat{C}^{(11)}$, 
and combinations thereof.
None of these operators connect states belonging to different symplectic 
representations and, furthermore, the SU(3) generators $\hat{C}^{(11)}_{1q}
=\hat{L}_q$ and $\hat{C}^{(11)}_{2\mu}=\frac{1}{\sqrt{3}}Q^E_{2\mu}$ act only 
within one level of excitation, $N$.  
Matrix elements for $\hat{C}^{(11)}$ in the standard SU(3) bases are given
by~\cite{JPD85,Hecht65}:
\bea
 \langle \lmp ||| \hat{C}^{(11)} ||| \lm \rangle &=&
   (-1)^{\phi_{\mu}} \sqrt{ 2 \langle \hat{C}_{SU3} \rangle [\lm]} 
   \; \delta_{\lmp \lm}  \; ,
\eea
where $\hat{C}_{SU3}$ denotes the second-order Casimir operator of SU(3),
given in Eq.~(\ref{eq:SU3Cas}), and $\phi_{\mu} = 1$ for $\mu \neq 0$ 
and $\phi_{\mu} = 0$ for $\mu = 0$.
The reduced matrix element $\langle \lmp ||| \hat{C}^{(11)} ||| \lm \rangle$
is related to the full  SU(3) matrix element via the Wigner-Eckart theorem
for SU(3) and the phase is chosen to be consistent with that of Ref.~\cite{JPD85}.

Several strategies for calculating matrix elements of the symplectic generators
$\hat{A}^{(20)}$ and $\hat{B}^{(02)}$ have been explored.
A direct way is to use the Sp(6,R) commutation relations to derive recursion
formulae, as shown by Rosensteel~\cite{Rosen80b}.
Another approach is to start from approximate matrix elements and to proceed
by successive approximations, adjusting the matrix elements until the
commutation relations are precisely satisfied~\cite{SymplM}.
Deenen and Quesne~\cite{Deenen84b} have employed a boson mapping to obtain 
generator matrix elements, and Casta\~{n}os {\em et al.}~\cite{Casta84b} have 
derived simple analytical functions for some special irreps.
The most elegant method, outlined by Rowe in Ref.~\cite{Rowe84b}, involves 
vector-valued coherent state representation theory and evaluates matrix 
elements of the symplectic raising and lowering operators by relating them 
to the matrix elements of a much simpler $u(3) \otimes \mbox{{\em Weyl}}$ 
algebra.
A listing of the relevant formulae is beyond the scope of this appendix, the
reader is thus referred to Ref.~\cite{Rowe84b} for details of the calculation.

Matrix elements of the SU(3) $\supset$ SO(3) integrity basis operators
$\hat{X}_3 \equiv (\hat{L} \times Q^E)_{(1)} \cdot \hat{L}$ and $\hat{X}_4 \equiv 
(\hat{L} \times Q^E)_{(1)} \cdot (\hat{L} \times Q^E)_{(1)}$ can be given in 
terms of SO(3) Racah recoupling coefficients 
$W(l_1,l_2,l,l_3;l_{12},l_{23})$~\cite{Var88} and the SU(3) $\supset$ SO(3) 
reduced matrix elements of $\hat{C}^{(11)}$~\cite{JPD85}:
\bea
\lefteqn{ \langle \lm \k l m | \hat{X}_3 | \lmp \k' l' m' \rangle } \nn \\
&=& 
\delta_{\lmp \lm} \delta_{l'l} \delta_{m'm} 3 l(l+1) \sqrt{2l+1}
\nn \\
&& \times W(l,1,l,1;l,2)
\langle \lm \k l || \hat{C}^{(11)}_2 || \lm \k' l \rangle ; 
\eea   
\bea
\lefteqn{ \langle \lm \k l m | \hat{X}_4 | \lmp \k' l' m' \rangle } \nn \\
&=& 
\delta_{\lmp \lm} \delta_{l'l} \delta_{m'm} 9 l(l+1) \sqrt{2l+1}
\nn \\
&& \times \sum_{\k'' l''} (-1)^{l+l''+1} \sqrt{2l''+1} \;\; [W(1,l,2,l'';l,1)]^2
\nn \\
&& \times \langle \lm \k l || \hat{C}^{(11)}_2 || \lm \k'' l'' \rangle 
\langle \lm \k'' l'' || \hat{C}^{(11)}_2 || \lm \k' l \rangle .  
\eea

\newpage

\begin{table}[hbtp]
\caption{B(E2) values (in Weisskopf units) for ground band transitions 
in $^{12}$C.
Compared are several symplectic calculations, PDS results, and experimental 
data.
Q denotes the static quadrupole moment of the $L^{\pi}=2^+_1$ state and is 
given in units of eb.  The experimental values are taken from
Refs.~\protect\cite{Ajz90,Vermeer83}.
PDS results are rescaled by an effective charge e$^*$=1.33 and
the symplectic calculations employ bare charges.}
\vspace{0.5cm}
\begin{center}
\begin{tabular}{ccccccccc}
\multicolumn{1}{c}{Transition} & \multicolumn{5}{c}{Model B(E2) [W.u.]} &
B(E2) [W.u.] \\
 \cline{2-6}
 $J_i \rightarrow J_f$ & \multicolumn{1}{c}{$2\hbar\om$} &
 \multicolumn{1}{c}{$4\hbar\om$} & \multicolumn{1}{c}{$6\hbar\om$} &
 \multicolumn{1}{c}{$8\hbar\om$} & \multicolumn{1}{c}{PDS} & Exp. \\
\hline
  2 $\rightarrow$ 0 & 4.65 & 4.65 & 4.65 & 4.65 & 4.65 &  4.65 $\pm$ 0.26\\
  4 $\rightarrow$ 2 & 4.35 & 4.27 & 4.24 & 4.23 & 4.28 &    n/a          \\
\hline
\multicolumn{1}{c}{
     $Q$ [eb]}      & 0.059& 0.060& 0.060& 0.060& 0.058& 0.06$\pm$ 0.03\\
\end{tabular}
\end{center}
\vspace{0.1in}
\label{BE2_C12}
\end{table}

\begin{table}[hbtp]
\caption{B(E2) values (in Weisskopf units)
for ground band transitions in $^{20}$Ne.
Compared are experimental data, predictions from several symplectic 
calculations, and PDS results.
The static quadrupole moment of the $L^{\pi}=2^+_1$ state is given in the 
last row.  
The experimental values are taken from 
Refs.~\protect\cite{Tilley98,Rag89,Spear81}.
PDS transition rates are rescaled by an effective charge e$^*$=1.95, while 
the symplectic calculations use bare charges.}
\begin{center}
\begin{tabular}{ccccccccc}
\multicolumn{1}{c}{Transition} & \multicolumn{5}{c}{Model B(E2) [W.u.]} &
B(E2) [W.u.] \\
 \cline{2-6}
 $J_i \rightarrow J_f$ & \multicolumn{1}{c}{$2\hbar\om$} &
 \multicolumn{1}{c}{$4\hbar\om$} & \multicolumn{1}{c}{$6\hbar\om$} &
 \multicolumn{1}{c}{$8\hbar\om$} & \multicolumn{1}{c}{PDS} & Exp. \\
\hline
  2 $\rightarrow$ 0 & 14.0 & 18.7 & 19.1 & 19.3 & 20.3 &  20.3 $\pm$ 1.0 \\
  4 $\rightarrow$ 2 & 18.4 & 24.5 & 24.6 & 24.5 & 25.7 &  22.0 $\pm$ 2.0 \\
  6 $\rightarrow$ 4 & 17.1 & 22.3 & 21.5 & 20.9 & 21.8 &  20.0 $\pm$ 3.0 \\
  8 $\rightarrow$ 6 & 12.4 & 15.2 & 13.3 & 12.4 & 12.9 &   9.0 $\pm$ 1.3 \\
\hline
\multicolumn{1}{c}{$Q$ [eb]}
    & -0.14 & -0.16 & -0.16 & -0.16 & -0.17 & -0.23 $\pm$ 0.03 \\
\end{tabular}
\end{center}
\vspace{0.1in}
\label{BE2_Ne20_GB}
\end{table}

\begin{table}[hbtp]
\caption{
Comparison of intraband and interband B(E2) rates for $^{20}$Ne.
Shown are various transitions between states of the lowest rotational bands.
K=$0_1$ denotes the ground band, which is dominated by 0$\hbo$ configurations;
members of the other bands correspond to 2$\hbo$ resonances.
Results are from the PDS calculation (rescaled by e$^*$=1.95) and from the 
8$\hbo$ symplectic approach (e$^*$=1.0).
In the last column, ratios of the calculated transition strengths are given.
}
\vspace{0.2cm}
\begin{center}
\begin{tabular}{ccccccc}
\multicolumn{4}{c}{Transition} & \multicolumn{2}{c}{Model B(E2) [W.u.]} &
\underline{BE2(PDS)} \\
\cline{5-6}
 $J_i$ & $K_i$ & $J_f$ & $K_f$ & Sp(6,R) & PDS & BE2(Sp6) \\
\hline
  2 & $0_1$ & 0 & $0_1$ & 19.3  & 20.3  & 1.05    \\
  2 & $0_2$ & 0 & $0_1$ &  5.8  & 12.6  & 2.16    \\
  2 & $0_3$ & 0 & $0_1$ &  0.10 &  0.32 & 3.16    \\
\hline
  2 & $0_1$ & 0 & $0_2$ &  2.9  &  5.7  & 1.94    \\
  2 & $0_2$ & 0 & $0_2$ & 20.3  & 27.8  & 1.37    \\
  2 & $0_3$ & 0 & $0_2$ &  0.15 &  0.13 & 0.84    \\
\hline
  2 & $0_1$ & 0 & $0_3$ &  0.17 &  0.48 & 2.80    \\
  2 & $0_2$ & 0 & $0_3$ &  0.25 &  0.26 & 1.01    \\
  2 & $0_3$ & 0 & $0_3$ & 12.9  & 16.8  & 1.30    \\
\hline
  4 & $0_1$ & 2 & $0_1$ & 24.5  & 25.7  & 1.05    \\
  4 & $0_2$ & 2 & $0_1$ & 10.9  & 22.8  & 2.09    \\
  4 & $1_1$ & 2 & $0_1$ &  2.3  &  5.8  & 2.55    \\
  4 & $2_1$ & 2 & $0_1$ &  0.63 &  2.3  & 3.66    \\
  4 & $0_3$ & 2 & $0_1$ &  0.09 &  0.30 & 3.34    \\
\hline
\end{tabular}
\end{center}
\vspace{0.1in}
\label{BE2_Ne20_Comp}
\end{table}

\begin{table}[hbtp] 
\caption{Dimensions of symplectic Hilbert spaces for $^{24}$Mg.
Shown are the number of $L$-states ($L=0,1,\ldots,8$) for spaces which
include $N\hbo$ excitations up to $N=0,2,4$, and 6, respectively.
}
\vspace{0.2cm}
\begin{center}
\begin{tabular}{crrrrrrrrr}
   Symplectic    & \multicolumn{9}{c}{Angular momentum $L$} \\
\cline{2-10}
     space       &   0 &   1 &   2 &   3 &   4 &   5 &   6 &   7 &   8 \\
\hline
     0$\hbo$     &   1 &   0 &   2 &   1 &   3 &   2 &   3 &   2 &   3 \\  
   (0+2)$\hbo$   &   4 &   3 &  11 &  10 &  17 &  15 &  19 &  16 &  18 \\  
  (0+2+4)$\hbo$  &  13 &  15 &  40 &  41 &  62 &  59 &  71 &  63 &  67 \\  
 (0+2+4+6)$\hbo$ &  32 &  49 & 110 & 122 & 171 & 171 & 198 & 182 & 187 \\          
\hline
\end{tabular}
\end{center}
\vspace{10.1in}
\label{Dim_Mg24}
\end{table}

\begin{table}[hbtp]
\caption{B(E2) strengths of ${}^{24}$Mg.
Compared are results from 2$\hbo$, 4$\hbo$, and 6$\hbo$ symplectic calculations,
a PDS calculation, and experiment~\protect\cite{Spear81,Branf75}.
Both intraband and interband transitions between states of the ground (K=$0_1$)
and $\gamma$ (K=$2_1$) band are given.
The static quadrupole moment of the $2^+_1$ state is listed in the last line 
(in units of eb)  
\protect\footnote{
Measurements have given results for $|Q|$ ranging from less than 0.16 eb to 
nearly double that value.  We list the value adopted in the review by 
Spear~\protect\cite{Spear81}.  }.   
The symplectic model reproduces the observed transition rates without employing
effective charges, while the PDS approach requires e$^*$=1.75.
}
\vspace{0.2cm}
\begin{center}
\begin{tabular}{ccccrrrrrrc}
\multicolumn{4}{c}{Transition} & \multicolumn{4}{c}{Model B(E2)} & B(E2) \\
\cline{5-8}
 $J_i$ & $K_i$ & $J_f$ & $K_f$ & \multicolumn{1}{c}{$2\hbo$} &
 \multicolumn{1}{c}{$4\hbo$} & \multicolumn{1}{c}{$6\hbo$} &
\multicolumn{1}{c}{PDS} &
Exp. \\
\hline
  2 & $0_1$ & 0 & $0_1$ & 17.2 & 20.2 & 20.4 & 20.5 & 20.5$\pm$0.6  \\
  4 & $0_1$ & 2 & $0_1$ & 24.5 & 26.9 & 26.9 & 26.2 &    23$\pm$4  \\
  6 & $0_1$ & 4 & $0_1$ & 25.2 & 25.5 & 25.2 & 22.5 &    34${}^{+36}_{-10}$  \\
  8 & $0_1$ & 6 & $0_1$ & 24.4 & 19.4 & 19.2 & 13.6 &    16${}^{+25}_{-6}$  \\
\hline
  3 & $2_1$ & 2 & $2_1$ & 31.6 & 35.6 & 35.3 & 36.6 &    34$\pm$6  \\
  4 & $2_1$ & 2 & $2_1$ &  9.7 & 11.2 & 11.0 & 11.6 &     16$\pm$3  \\
  5 & $2_1$ & 3 & $2_1$ & 15.3 & 17.0 & 16.6 & 16.8 &    28$\pm$5  \\
  5 & $2_1$ & 4 & $2_1$ & 17.3 & 18.0 & 17.7 & 18.0 &    14$\pm$6  \\
  6 & $2_1$ & 4 & $2_1$ & 15.3 & 19.4 & 18.3 & 20.1 &    23${}^{+23}_{-8}$  \\
  8 & $2_1$ & 6 & $2_1$ & 12.4 & 18.0 & 15.9 & 19.6 &       $\geq$3  \\
\hline
  2 & $2_1$ & 0 & $0_1$ &  1.1 &  1.3 &  1.3 &  3.1 &   1.4$\pm$0.3  \\
  2 & $2_1$ & 2 & $0_1$ &  2.2 &  1.7 &  1.9 &  3.4 &   2.7$\pm$0.4  \\
  3 & $2_1$ & 2 & $0_1$ &  1.9 &  2.4 &  2.3 &  5.6 &   2.1$\pm$0.3  \\
  4 & $2_1$ & 2 & $0_1$ &  0.2 &  1.0 &  0.9 &  2.7 &   1.0$\pm$0.2  \\
  4 & $2_1$ & 4 & $0_1$ &  2.9 &  2.1 &  2.3 &  4.1 &   1.0$\pm$1.0  \\
  5 & $2_1$ & 4 & $0_1$ &  1.0 &  2.4 &  2.0 &  6.0 &   3.9$\pm$0.8  \\
  6 & $2_1$ & 4 & $0_1$ &  0.2 &  1.2 &  1.0 &  3.2 & 0.8${}^{+0.8}_{-0.3}$  \\
\hline
\multicolumn{4}{c}{$Q$ [eb]}
                & -0.171 & -0.186 & -0.185 & -0.191 & -0.18$\pm$0.02  \\
\end{tabular}
\label{BE2_Mg24_GG}
\end{center}
\end{table}

\begin{table}[phbt]
\caption{Comparison of intraband and interband B(E2) rates for $^{24}$Mg.
Shown are selected transitions between states of the K=$0_1,0_2,0_3$,
and $2_2$ bands.
The PDS values are rescaled by e$^*$=1.75.
Ratios of the results from the two theoretical approaches are given in
the last column.
} 
\vspace{0.2cm}
\begin{center}
\begin{tabular}{ccccccc}
\multicolumn{4}{c}{Transition} & \multicolumn{2}{c}{Model B(E2) [W.u.]} &
\underline{BE2(PDS)} \\
\cline{5-6}
 $J_i$ & $K_i$ & $J_f$ & $K_f$ & Sp(6,R) & PDS & BE2(Sp6) \\
\hline
  2 & $0_1$ & 0 & $0_1$ & 20.4  & 20.5 & 1.00    \\
  2 & $0_2$ & 0 & $0_1$ &  5.6  & 10.2 & 1.84    \\
  2 & $0_3$ & 0 & $0_1$ &  0.047&  0.19& 4.09    \\
  2 & $2_2$ & 0 & $0_1$ &  0.22 &  2.1 & 9.46    \\ 
\hline
  2 & $0_1$ & 0 & $0_2$ &  2.5  &  5.2 & 2.05    \\
  2 & $0_2$ & 0 & $0_2$ & 14.8  & 26.6 & 1.80    \\
  2 & $0_3$ & 0 & $0_2$ &  0.037& 0.047& 1.26    \\
  2 & $2_2$ & 0 & $0_2$ &  0.48 &  3.4 & 7.00    \\  
\hline
  2 & $0_1$ & 0 & $0_3$ &  0.025& 0.042 & 1.69    \\
  2 & $0_2$ & 0 & $0_3$ &  0.12 & 0.12  & 1.06    \\
  2 & $0_3$ & 0 & $0_3$ & 12.9  & 16.2  & 1.26    \\
  2 & $2_2$ & 0 & $0_3$ &  0.023&  0.12 & 5.21    \\  
\hline
  4 & $0_1$ & 2 & $0_1$ & 26.9  & 26.2  & 0.97    \\
  4 & $0_2$ & 2 & $0_1$ &  9.7  & 18.4  & 1.90    \\
  4 & $0_3$ & 2 & $0_1$ &  0.052&  0.48 & 9.20    \\
  4 & $2_2$ & 2 & $0_1$ &  0.66 &  0.21 & 0.32    \\  
\hline
\end{tabular}
\end{center}
\vspace{0.1in}
\label{BE2_Mg24_Comp}
\vskip 3cm 
\end{table}

\newpage

\end{document}